\documentclass[twocolumn,pra,showpacs,superscriptaddress,showpacs]{revtex4-1}
\usepackage{amssymb,amsmath}
\usepackage{graphicx}
\usepackage{bm}
\usepackage{dcolumn}
\usepackage{float}
\usepackage{url}
\usepackage{color}
\usepackage{subfigure}
\usepackage{txfontsb}
\usepackage[OT1]{fontenc} 
\usepackage[driverfallback=dvipdfm]{hyperref}
\hypersetup{pdfpagemode=FullScreen,colorlinks=true,breaklinks,urlcolor=blue,linkcolor=blue,citecolor=blue}

\usepackage{braket}
\usepackage{enumerate}
\usepackage{xcolor}

\begin{document}

\title{Linking invariant for the quench dynamics of a two-dimensional two-band Chern insulator}

\author{Xin Chen}
\affiliation{Institute for Advanced Study, Tsinghua University, Beijing 100084, China}

\author{Ce Wang}
\affiliation{Institute for Advanced Study, Tsinghua University, Beijing 100084, China}

\author{Jinlong Yu}
\email{jinlong.yu.physics@gmail.com}
\affiliation{Center for Quantum Physics, University of Innsbruck, Innsbruck A-6020, Austria}
\affiliation{Institute for Quantum Optics and Quantum Information of the Austrian Academy of Sciences, Innsbruck A-6020, Austria}
\affiliation{Institute for Advanced Study, Tsinghua University, Beijing 100084, China}

\date{\today}

\begin{abstract}
We discuss the topological invariant in the (2+1)-dimensional quench dynamics of a two-dimensional two-band Chern insulator starting from a topological initial state (i.e., with a nonzero Chern number $c_i$), evolved by a post-quench Hamiltonian (with Chern number $c_f$). This process is classified by the \emph{torus} homotopy group $\tau_3(S^2)$. In contrast to the process with $c_i=0$ studied in previous works, this process cannot be characterized by the Hopf invariant that is described by the sphere homotopy group $\pi_3(S^2)=\mathbb{Z}$. It is possible, however, to calculate a variant of the Chern-Simons integral with a complementary part to cancel the Chern number of the initial spin configuration, which at the same time does not affect the (2+1)-dimensional topology. We show that the modified Chern-Simons integral gives rise to a topological invariant of this quench process, i.e., the linking invariant in the $\mathbb{Z}_{2c_i}$ class: $\nu = (c_f - c_i) \mod (2c_i)$. We give concrete examples to illustrate this result and also show the detailed deduction to get this linking invariant.  
\end{abstract}

\maketitle
\section{Introduction}

Two-dimensional (2D) two-band Chern insulators have been among the most basic and active research subjects in condensed matter physics. In a seminal work by Haldane~\cite{Haldane1988}, he proposed a 2D two-band model describing spinless fermions hopping on a hexagonal lattice in the presence of a staggered magnetic field. This system features an integer-valued topological number---the Chern number~\cite{TKNN1982}, which gives rise to quantized Hall response. With the increasing controllability of ultracold atoms as powerful quantum simulators, the Haldane model~\cite{Haldane1988}, and its close cousin---the quantum anomalous Hall model on a square lattice~\cite{Qi2006}, have been realized in recent ultracold atom experiments~\cite{TopHaldane,Wu2016}. 

In addition to studying the ground state of the Chern insulator, the ultracold atom system is also suited to investigate out-of-equilibrium properties therein~\cite{Caio2015,Caio2016,Hu-Zoller2016,Unal2016,Wilson2016,Ce,Yu2017,Chang2018,Ezawa,Yi2018A,Liu2018,Yu2019Hopf,Zhang-Zhang-Liu2019_Dyn-Top-Charges,  Liu-Int-quench2019,Liu-shallow-quench2019,Unal-Eckardt2019,Flaschner2018,chernlink,USTC-PKU2018,Hopf_fibration2019,Yi-Shuai2019}; see also related works on quantum dynamics for one-dimensional systems~\cite{Gong-Ueda2018,Yang-Chen2018,Cooper2018,YiWei2019,Lu2019}.  In several recent experiments, the Hamburg group~\cite{Flaschner2018,chernlink} and the PKU-USTC group~\cite{USTC-PKU2018,Yi-Shuai2019,Hopf_fibration2019} have studied the quench process of a Chern insulator, starting with a topologically \emph{trivial} initial state and evolved by a post-quench Hamiltonian. As shown in Ref.~\cite{Ce}, it is possible to define a topological invariant---the Hopf invariant, which is an integration of Chern-Simons density over two-dimensional quasimomenta and one-dimensional time---for this (2+1)-dimensional quench dynamics. It is also proved in Ref.~\cite{Ce} that this dynamical Hopf invariant exactly equals the Chern number of the post-quench Hamiltonian. The geometric essence of the Hopf invariant is the linking invariance (or, the linking number), which gives rise to the structure of Hopf fibration in the quench dynamics~\cite{Yu2019Hopf,Hopf_fibration2019}. Here, a fiber is a trajectory in the (2+1)-dimensional quasimomentum-time space $T^3$ whose spin vector points to a given direction on the Bloch sphere $S^2$. The linking invariance tells how many times that any two fibers are linked. And it has been extracted experimentally by identifying the fibers for the North and South Poles of $S^2$ experimentally~\cite{chernlink,USTC-PKU2018,Hopf_fibration2019}. 

Almost all of the aforementioned studies on the quench dynamics of Chern insulators focus on the quench process starting with a topologically trivial initial state. A natural question then arises: What is the topological invariant for the quench dynamics starting with a \emph{topological} initial state? Surprisingly, it turns out that this question has been addressed more than three decades ago by Wilczek and Zee in a different context~\cite{Azee}. They have studied the motion of a two-dimensional Skyrmion (with Chern number $c_i = -1$) evolved in a constant magnetic field (with Chern number $c_f=0$), and defined the Hopf invariant ($\nu=1$ as they claim) as the integration of the Chern-Simons density for the quench dynamics of the Skyrmionic spin pattern~\cite{Azee}. Indeed, this work inspired our previous theoretical investigation of the quench dynamics starting with a topologically trivial initial state~\cite{Ce}. However, it turns out to be an inappropriate way to define the topological invariant for the quench process with a \emph{topological} initial state, because both the Chern-Simons density and its integration are gauge dependent; see Eq.~\eqref{eq:constH} below. Hence, we show in Eq.~\eqref{eq:constH} that  {\lq\lq}the Hopf invariant{\rq\rq} defined this way is \emph{not} an invariant at all. The key problem for identifying the Chern-Simons integral as a topological invariant is the following: As the initial state has nonzero Chern number, it puts an obstruction to globally define a continuous gauge for the Chern-Simons density to evaluate the Chern-Simons integral~\cite{Teo-Kane2010}. We notice that, there is a recent work by Ezawa~\cite{Ezawa}, who has also considered the quench process starting with a topological initial state. In Ref.~\cite{Ezawa}, the author also makes use of the Chern-Simons integral as the definition of the topological invariant (as in Ref.~\cite{Azee}) for the quench process. Such a treatment is inappropriate.  

Although the Chern-Simons integral fails to be a topological invariant describing the quench process with a topological initial state, we can still make use of a certain variant of it to extract the genuine (2+1)-dimensional topological invariant, i.e., the linking invariant, for this quench process. The key of our recipe is to introduce a complementary part in the two-dimensional quasimomentum space, as detailed in Sec.~\ref{Sec:CS_To_Linking}, to cancel the Chern number of the initial spin configuration. The magnetic field in this complementary part is chosen to be either parallel  or anti-parallel to the spins therein, and thus the introduction of this part does not lead to any change of the (2+1)-dimensional topology: It does not give rise to any nontrivial linking, and the linking invariant remains intact. As the Chern number for the initial spin configuration is nullified to be zero when considering the system with the complementary part, we can use the Chern-Simons integral to evaluate the topological invariant in this modified quasimomentum-time manifold for this quench process. 

This paper is organized as follows. In Sec.~\ref{Sec:MainResults}, we introduce the quench process that we study, and give the expression of the topological invariant in this quench process supplemented with concrete examples to validate it. In Sec.~\ref{Sec:CalculateTopInv}, we show the details about the calculation of the linking invariant for the quench process. We first show why there is a problem to directly use the Chern-Simons integral to evaluate the quench from a topological initial state, and then show how this problem is solved by introducing a complementary part to nullify the topology of the initial spin configuration. We also show that the linking invariant is defined only module $2c_i$. In Sec.~\ref{Sec:Experimental-feasibility}, we discuss briefly the experimental feasibility of our protocol, focusing on the possibility of preparing a topological initial state in ultracold quantum gases~\cite{Simone-Jinlong-Peter-Jan2019}. We conclude in Sec.~\ref{Sec:Conclusion}.

\section{Main results with examples} \label{Sec:MainResults}

\subsection{Quench dynamics for a 2D two-band model}
 We work on a general two-band tight-binding model in two dimensions with the following form:
\begin{align}
{\cal H}(\mathbf{k})=\frac{1}{2}\mathbf{H}(\mathbf{k})\cdot \boldsymbol{\sigma},\label{eq:Ham}
\end{align} 
where $\mathbf{k}=(k_x, k_y)$ is the quasimomentum, and $\boldsymbol{\sigma}=(\sigma_x,\sigma_y,\sigma_z)$ is the vector of Pauli matrices. The evolution time $t$ can be rescaled as $t\rightarrow {\tilde t = |\mathbf{H}|t}$, so that at $\tilde t={2\pi}$ every spin returns to its initial position. The rescaling is intended to make the calculations simple, since the topological classification does not depend on the energy dispersion as long as the band gap is non-vanishing. The rescaling of time $t$ is equivalent to rescaling of the Hamiltonian vector $\mathbf{H}(\mathbf{k})$. In the following, we use $\mathbf{h}(\mathbf{k})=\mathbf{H}(\mathbf{k})/|\mathbf{H}(\mathbf{k})|$ to denote the rescaled (i.e., normalized) Hamiltonian vector. 

We consider a half-filled insulator state $\ket{\xi_0(\mathbf{k})}$ as the initial state, which is the ground state of a topological Hamiltonian with $\mathbf{h}(\mathbf{k})=\mathbf{h}^i(\mathbf{k})$. This state is evolved under another Hamiltonian with $\mathbf{h}(\mathbf{k})=\mathbf{h}^f(\mathbf{k})$:
\begin{equation}
	\ket{\xi(\mathbf{k},t)} = U_t(\mathbf{k},t) \ket{\xi_0(\mathbf{k})},
\end{equation}
where 
\begin{equation}
 U_t(\mathbf{k},t)=\exp\left(-\frac{i}{2}\mathbf{h}^f(\mathbf{k})\cdot \boldsymbol{\sigma} t\right)
\end{equation}
is the evolution operator. In the expression above, the evolution time should be understood as the rescaled time ${\tilde t = |\mathbf{H}|t}$, and the tilde of $\tilde t$ has been dropped for simplicity (and without confusion). 
The quasimomentum- and time-dependent Bloch vector is written as
\begin{equation}\label{Eq:n_k_t}
\mathbf{n}(\mathbf{k},t)=\bra{\xi(\mathbf{k},t)}\bm{\sigma}\ket{\xi(\mathbf{k},t)}.
\end{equation}
Equation \eqref{Eq:n_k_t} forms a natural mapping from a three-torus $(k_x,k_y,t)\in T^3$ to a two-sphere $S^2$ (i.e., the Bloch sphere), which is generally characterized by the torus homotopy group $\tau_3(S^2)$~\cite{Fox1948}; more details on this point will be elaborated below. The two-sphere could be parametrized by the spherical coordinate $(\theta, \phi)$: ${\mathbf{n}} = ({n_x},{n_y},{n_z}) = (\sin \theta \cos \phi ,\sin \theta \sin \phi ,\cos \theta )$. 
For one point on $S^2$, its inverse image is a one-dimensional manifold (which is generally not contractible) in $T^3$. 

\subsection{Topological invariant in the quench dynamics: $\mathbb{Z}_{2c_i}$ linking number}
In contrast to the Hopf mapping of $\pi_3(S^2)=\mathbb{Z}$, which describes the dynamical process of the system with nearly-polarized trivial initial spin configuration~\cite{Ce}, we find out that the quench process with a topological initial spin configuration is classified by a doublet $(c_i; \nu)$, where $\nu\in \mathbb{Z}_{2c_i}$ is the linking invariant. The expression for $\nu$ is as follows:
\begin{align}
\nu = (c_f-c_i)\mod (2c_i),\label{eq:mainrslt}
\end{align}
where $c_i$ is the Chern number of  the initial spin configuration, and $c_f$ is the Chern number of the post-quench Hamiltonian. Note that, in mathematics this process is classified by the torus homotopy group $\tau_3(S^2)$~\cite{Fox1948}. The torus homotopy group $\tau_3(S^2)$ could be characterized by a quadruplet $(c_x,c_y,c_i;\nu)$, where $(c_x,c_y,c_i)$ are the Chern numbers of the three 2-dimensional tori $\{T_{yt}(k_x^{(0)},k_y,t),T_{xt}(k_x, k_y^{(0)},t),T_{xy}(k_x,k_y,t^{(0)})\}$, and $\nu$ is the Pontryagin invariant~\cite{Pontryagin1941}. In the notation for the tori, e.g., $T_{yt}(k_x^{(0)},k_y,t)$, the $k_x$ index takes a given constant value $k_x^{(0)}$, while the variable $k_y$ takes value in the first Brillouin zone and $t\in[0,2\pi)$. Such two-dimensional tori are different sections of $T^3$. The Pontryagin invariant $\nu$ satisfies the following equivalence relationship~\cite{Pontryagin1941}:
\begin{align}
\nu^\prime \equiv \nu \mod 2\text{gcd}(c_{x},c_{y},c_{i}), \label{eq:THG}
\end{align}
where the function $\text{gcd}$ evaluates the greatest common divisor of its variables (which are integers). 
It could be shown that, in the dynamical quench process with a topologically nontrivial initial spin configuration, $c_{x}=c_{y}=0$ (see Appendix \ref{App.A} for details). We then see that $c_i = \text{gcd}(c_x=0,c_y=0,c_i)$, and Eq.~\eqref{eq:THG} is consistent with Eq.~\eqref{eq:mainrslt}. For the case of $c_i=0$, this classification reduces to $\nu=c_f\in\mathbb{Z}$ which is consistent with the result of Hopf mapping of the conventional (sphere) homotopy group $\pi_3(S^2)=\mathbb{Z}$ as studied in the previous works~\cite{Ce,Yu2019Hopf}. 

The way to detect the linking invariant $\nu$ is by counting the twisting times (i.e., the linking number) of two loops in $T^3$, where 
each loop is constitute of the preimage of one point on $S^2$ and another auxiliary straight lines in the time direction (these auxiliary lines will be explained later). 
The classification above has the following consequences. 
First, for the quench processes with same $c_i$ but different $c_f$ values that differ by $2c_i$, the corresponding linking numbers may be the same [cf. Figs. \ref{Fig2}(a) and \ref{Fig2}(d) below]. Second, for the same quench process, the linking number of two loops may differs by $2c_i$ due to the different choices of complementary parts and hence, different auxiliary lines [cf. Figs. \ref{Fig2}(a) and \ref{Fig2}(b) below]. 

\subsection{Examples}

\begin{figure}[btp]
\includegraphics[width=\columnwidth]{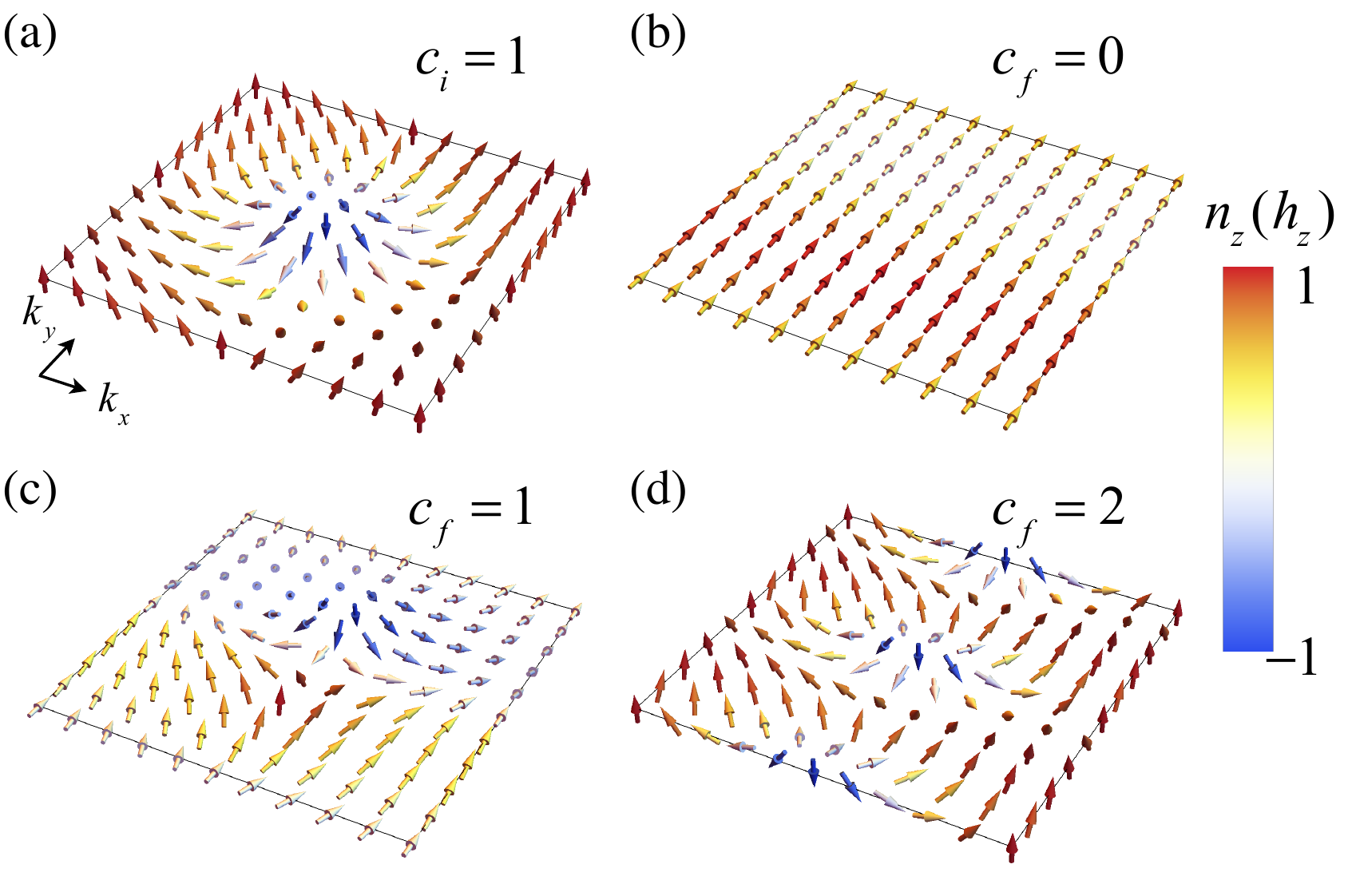}
\caption{\label{Fig1} The distribution of the Bloch vectors $\mathbf{n}_i(\mathbf{k})$ for the initial state with Chern number $c_i = 1$ (a), and the vectors of the post-quench Hamiltonians $\mathbf{h}^f(\mathbf{k})$ with different Chern numbers $c_f$ (b-d). The squares in the figures represent the first Brillouin zone $\{ ({k_x},{k_y})|{k_x} \in [ - \pi ,\pi ),{k_y} \in [ - \pi ,\pi )\} $. The arrows are colored by the $z$ components $n_z$ (or $h_z$) of the corresponding vectors $\mathbf{n}$ (or $\mathbf{h}$).}
\end{figure}

We use some concrete examples to illustrate the result shown in Eq.~\eqref{eq:mainrslt}. In particular, we take the initial state as a topological ground state of the Hamiltonian ${\cal H}(\mathbf{k})=\frac{1}{2}\mathbf{H}(\mathbf{k})\cdot \boldsymbol{\sigma}$ with~\cite{Qi2006}
\begin{eqnarray}\label{Eq:Hami_QAH}
\begin{aligned}
H_x(\mathbf{k})&=\sin(k_x),\\
H_y(\mathbf{k})&=\sin(k_y),\\
H_z(\mathbf{k})&=2-M-\cos(k_x)-\cos(k_y).
\end{aligned}\label{eq:skr}
\end{eqnarray}
The Chern number of this Hamiltonian as a function of $M$ is 
\begin{equation}\label{Eq:c_M_1}
	c(M) = \left\{ \begin{aligned}
  1&,\quad \quad &{\text{for}}\quad&0 < M < 2; \hfill \\
   -1&,\quad\quad &{\text{for}}\quad&2 < M < 4; \hfill \\
  0&,\quad\quad &{\text{for}}\quad&|M-2|>2. \hfill \\ 
\end{aligned}  \right.
\end{equation}
We take $M_i=1$, and thus the Chern number of the initial state is $c_i=1$. For this case, the Bloch vector $\mathbf{n}_i(\mathbf{k})=\mathbf{n}(\mathbf{k},t=0)$ in the first Brillouin zone is shown in Fig.~\ref{Fig1}(a). 

We consider three types of post-quench Hamiltonians, whose normalized Hamiltonian vectors are shown in Figs.~\ref{Fig1}(b)-\ref{Fig1}(d). The first one is a topologically trivial Hamiltonian with Chern number $c_f = 0$~[Fig.~\ref{Fig1}(b)]. 
For this case, we choose the evolving Hamiltonian with the form of Eq.~\eqref{Eq:Hami_QAH} with $M_f=15.5$, but the spin basis is rotated around the $x$ axis for $90^\circ$~\cite{Note_rotate_x}. 
The second one is an evolving Hamiltonian with Chern number $c_f=1$ [Fig.~\ref{Fig1}(c)]. For this case, we choose the evolving Hamiltonian with the form of Eq.~\eqref{Eq:Hami_QAH} with $M_f=0.5$, but the spin basis is also rotated around the $x$ axis for $90^\circ$. 
The third one is a Hamiltonian with Chern number $c_f=2$ [Fig.~\ref{Fig1}(d)]. To realize this, we use the post-quench Hamiltonian ${\cal H}(\mathbf{k})=\frac{1}{2}\mathbf{H}(\mathbf{k})\cdot \boldsymbol{\sigma}$ with
\begin{eqnarray}
\begin{aligned}
H_x(\mathbf{k})&=\sin(k_x),\\
H_y(\mathbf{k})&=\sin(2k_y),\\
H_z(\mathbf{k})&=2-M-\cos(k_x)-\cos(2k_y).
\end{aligned}
\end{eqnarray}
Because of the factor $2$ associated with the quasimomentum $k_y$, the Chern number of this Hamiltonian as a function of $M$ is doubled compared with Eq.~\eqref{Eq:c_M_1}:
\begin{equation}\label{Eq:c_M_2}
	c(M) = \left\{ \begin{aligned}
  2&,\quad \quad &{\text{for}}\quad&0 < M < 2; \hfill \\
   -2&,\quad\quad &{\text{for}}\quad&2 < M < 4; \hfill \\
  0&,\quad\quad &{\text{for}}\quad&|M-2|>2. \hfill \\ 
\end{aligned}  \right.
\end{equation}
We take $M_f=1$ for this post-quench Hamiltonian, and the corresponding Hamiltonian vector is shown in Fig.~\ref{Fig1}(d). 

\begin{figure}[tbp]
\includegraphics[width=\columnwidth]{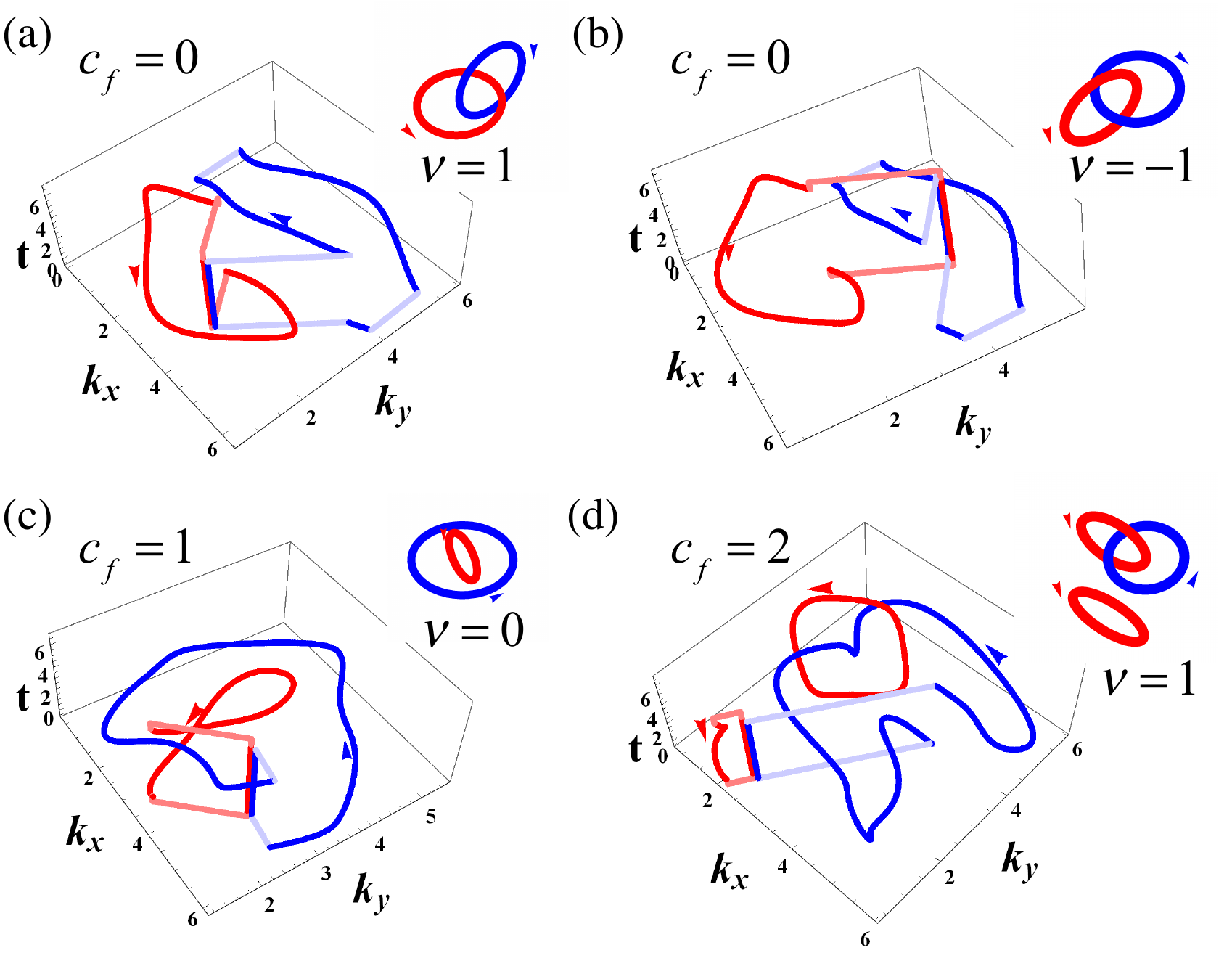}
\caption{\label{Fig2}
Extracting the linking numbers for different quench processes starting with a topological initial state with $c_i=1$, evolved by different post-quench Hamiltonians with $c_f=0$ (a, b), $c_f=1$ (c) and $c_f=2$ (d).  
The figures show the preimage loops of two points on the Bloch sphere with $(\theta,\phi)=(\pi/4,\pi/5)$ and $(3\pi/4,6\pi/5)$ (the antipodal point of the former). The straight lines going along the time direction are the auxiliary lines used to define the linking of two loops. Some other auxiliary lines in lighter colors  are added to make the separated lines connected (to form closed loops). Such auxiliary lines can cancel in pairs because of the periodicity of the torus (note that their directions are always opposite). The post-quench Hamiltonians are the same for (a) and (b), and the only difference between them is the different choice the positions of the auxiliary straight lines along the time direction. To facilitate the identification of the linking numbers, we have also made schematic illustrations on the top right of each figures, which show the circles that are homeomorphic to the corresponding ones in the main figures. The sign of the linking number is determined by the right-hand rule.}
\end{figure}

Figure \ref{Fig2} shows the preimages of two points on the Bloch sphere $S^2$ for the quench processes with the Chern number of the initial spin configuration $c_i=1$. 
It is illustrated from the figure that by adding an auxiliary straight line to a preimage loop, the loop forms the boundary of a two-dimensional surface. And then the notion of linking number makes sense. In Fig. \ref{Fig2}, the linking number of $c_f=0$ is the same as $c_f=2$, this is an evidence that the classification of this dynamical process for $c_i=1$ is well defined modulo $2$. Fig. \ref{Fig2}(a) and Fig. \ref{Fig2}(b) are the twisting loops of two different pairs in the same quench process. The difference of Fig. \ref{Fig2}(a) and Fig. \ref{Fig2}(b) is the choice the position of the auxiliary straight line. Their linking numbers differ by $2$. 

\section{Calculation of the topological invariant} \label{Sec:CalculateTopInv}
\subsection{Problem of the Chern-Simons integral with a topological initial state}
We first define a Chern-Simons integral as follows~\cite{Azee,Moore-Wen2008}:
\begin{equation}\label{Eq:I_CS}
{I_{{\text{CS}}}} = \frac{1}{{4{\pi ^2}}}\int_{{\text{BZ}}} {d{k_x}d{k_y}} \int_0^{2\pi } {dt} {\mathcal{A}_\mu }{\mathcal{J}_\mu },
\end{equation}
where the index $\mu$ takes values in $k_x$, $k_y$, and $t$, and Einstein's summation convention for the indexes is used throughout. 
Here, ${\cal A}_\mu=-i \langle\xi({\bf k},t)|\partial_{\mu}|\xi({\bf k},t)\rangle$ is the Berry connection, with $|\xi({\bf k},t)\rangle$ being the time-dependent Bloch state $|\xi({\bf k},t)\rangle=U_t({\bf k})|\xi_0(\bf{k})\rangle$. ${\mathcal{J}_\mu } = {\epsilon ^{\mu \nu \rho }}{\partial _\nu }{\mathcal{A}_\rho }$ is the Berry curvature, where $\epsilon^{\mu\nu\rho}$ is the 
3-dimensional Levi-Civita symbol.
The integrand of the integral above can be rewritten as ${\mathcal{A}_\mu }{\mathcal{J}_\mu } = {\epsilon ^{\mu \nu \rho }}{\mathcal{A}_\mu }{\partial _\nu }{\mathcal{A}_\rho }$, which takes the form of a Chern-Simons term (see, e.g., \cite{Baez1994_Gauge_Knots}). The domain of integration for the quasimomentum is the first Brillouin zone (BZ for short), which is a 2-torus $T^2$. And this Chern-Simons integral is defined on the 3-torus $T^3 = T^2\times S^1$, with $S^1 = [0,2\pi)$. The Berry curvature, which is also the tangent vector of the preimage loop, could be simplified in two band systems~\cite{Qi2006,Azee}:
\begin{align}
{\cal J}_{\mu}&=\frac{1}{4}{\epsilon ^{\mu \nu \rho }}\mathbf{n}\cdot(\partial_\nu \mathbf{n}\times \partial_\rho \mathbf{n}). 
\end{align}
It is a gauge independent quantity. 

For the quench process with $c_i=0$, the Chern-Simons integral is a topological invariant, termed as the Hopf invariant, which equals the Chern number of the post-quench Hamiltonian $c_f$~\cite{Ce}. However, for the quench process with $c_i\neq0$, it is no longer valid to identify this integral as a topological invariant for the $(2+1)$-dimensional quench dynamics, because it is a gauge-dependent and, generally, \emph{not quantised} quantity for this case. 
As an example, we reconsider the case that a Skyrmionic spin pattern (with Chern number $c_i = 1$) rotates within a constant magnetic field (which can be regarded as a constant post-quench Hamiltonian with Chern number $c_f = 0$) as discussed in Ref.~\cite{Azee}. The Chern-Simons integral for this case is given by (see Appendix \ref{sec:CSGauge} for details)
\begin{align}
I_\text{CS} = -\frac{1}{2} \mathbf{d}\cdot \mathbf{h}^f. \label{eq:constH}
\end{align}
Here, $\mathbf{h}^f$ is the direction of the constant magnetic field and $\mathbf{d}$ is related to the gauge fixing: The gauge is fixed such that the connection ${\cal A}_\mu$ is discontinuous at $\mathbf{n}(k_x,k_y,t) = \mathbf{d}$, i.e., $\mathbf{d}$ is the direction of the Dirac string. Equation \eqref{eq:constH} shows that the Chern-Simons integral depends on the gauge, even for the constant magnetic field case. In the calculation of \cite{Azee}, the gauge is (implicitly) chosen as $\mathbf{d}=-\mathbf{h}^f$ so that the integral in this gauge gives rise to $I_\text{CS}= \frac{1}{2}$, which is still not an integer. 
The failure of getting a topological invariant from the Chern-Simons integral is rooted in the \emph{weak} two-dimensional topology, i.e., the nontrivial initial Chern number $c_i$, in the (2+1)-dimensional quench process. This nonvanishing $c_i$ puts an obstruction to globally define a continuous gauge for the Berry connection $\mathcal{A}_\mu$ (and hence, the Chern-Simons density ${\mathcal{A}_\mu }{\mathcal{J}_\mu }$) to evaluate the Chern-Simons integral~\cite{Teo-Kane2010}. 

The Chern-Simons integral actually calculates how many times one loop (which is a one-dimensional curve in $T^3$ that map to a particular point on the Bloch sphere $S^2$) crosses a surface bounded by another loop. The linking number is also defined as the times one loop crosses a surface bounded by another loop. From this point, it seems contradictory for the $c_i\neq0$ case that the linking number should be gauge independent but the Chern-Simons integral is gauge dependent. 
The key to resolving this apparent contradiction is as follows. 
For the case of topologically trivial spin configuration (with $c_i=0$), every preimage loop could shrink to one point and then the surface bounded by this loop is well-defined. However, in the topological nontrivial case with $c_i\neq0$, the preimage loops are the loops winding $c_i$ times along the time direction circle. Therefore, a preimage loop fail to shrink to one point, and then cannot be the boundary of any surface. 
For this case, the linking number is not even well defined, which is consistent with the fact that the Chern-Simons integral is gauge dependent. Thus we see that the Chern-Simons integral in its current form can not be identified as the linking number between topologically nontrivial preimage loops. 

\subsection{Chern-Simons integral with a complementary part: cancelation of the weak 2D topology} \label{Sec:CS_To_Linking}
Nevertheless, we could still use a variant of the Chern-Simons integral to obtain the linking number by clinging a complementary patch to the original Brillouin zone. 
In order to make the notion of linking number well-defined, we combine an auxiliary straight line in the time direction with the original one, so that the two loops together is topologically trivial in fundamental group: They form the boundary of a certain surface (cf., Fig.~\ref{Fig2}). The boundary of this surface can be viewed as the preimage loop of a point on $S^2$ in a dynamical process with total initial Chern number $c_i^{(\text{tot})} = 0$ (as elaborated below). 

\begin{figure}[tbp]
\includegraphics[width=\columnwidth]{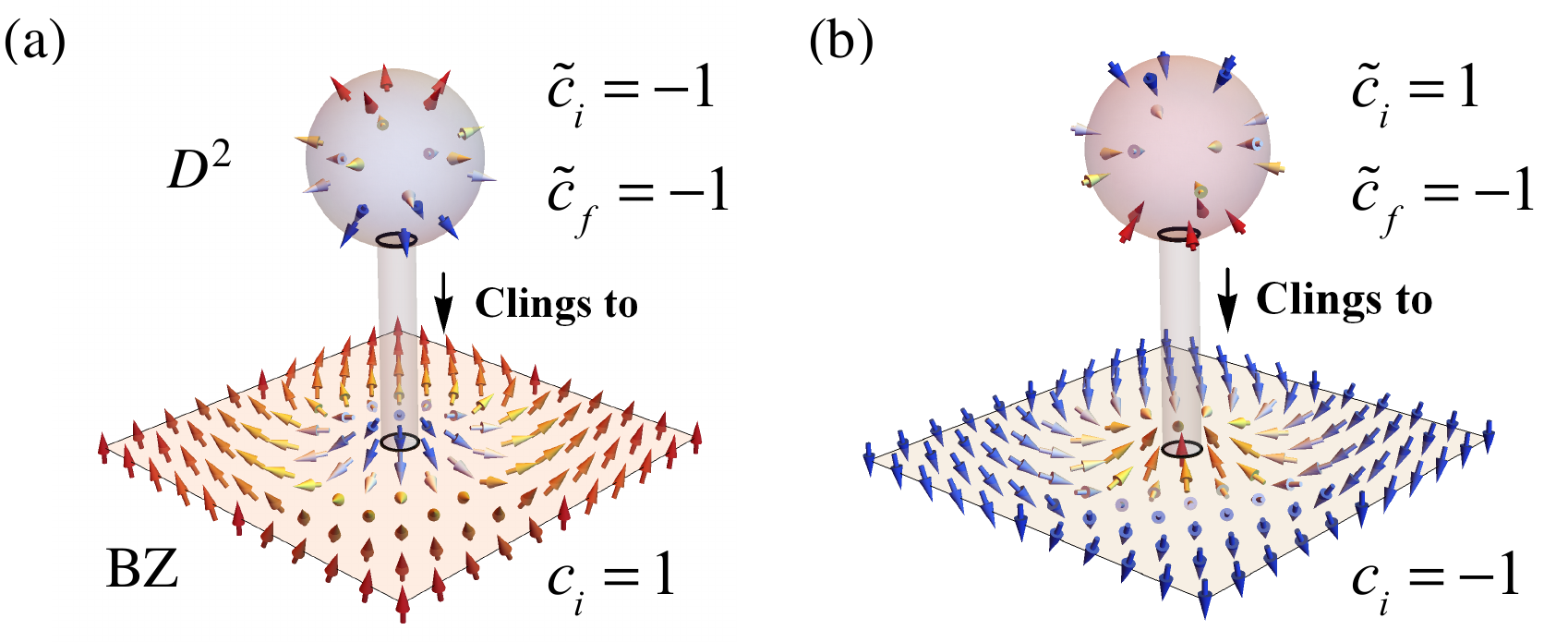}
\caption{\label{Fig3}
Illustration of the operation of clinging a $D^2$ to the Brillouin zone (which is a 2-torus denoted as $T^2$) at a quasimomentum point $\bf{s}$. The arrows indicate the spin directions ($\mathbf{n}$ on $T^2$ and $\tilde{\mathbf{n}}$ on $D^2$). We assume the magnetic field $\tilde{\mathbf{h}}$ on the sphere is consistent with the magnetic field $\mathbf{h}$ on the Brillouin zone. We further assume the magnetic field for the sphere is always pointing out along the normal vector direction of the sphere, which always gives Chern number $\tilde{c}_f=-1$. Then (a) shows the case that $\mathbf{n}(\mathbf{s},t=0)$ is parallel to $\mathbf{h}(\mathbf{s})$, and (b) shows the case that $\mathbf{n}(\mathbf{s},t=0)$ is anti-parallel to $\mathbf{h}(\mathbf{s})$. In both cases, the total Chern numbers for the initial spin configurations on the modified Brillouin zone ${{\tilde T}^2} = ({T^2}\backslash {\mathbf{s}}) \cup {D^2}$ (which is still a 2-torus) are zero.}
\end{figure}

Adding the auxiliary line (as done in Fig.~\ref{Fig2}) is equivalent to clinging a complementary patch to the original Brillouin zone (see Fig.~\ref{Fig3}) with a given monopole-like spin (with Chern number $\tilde{c}_i=-c_i$) and magnetic field (with Chern number $\tilde{c}_f = \pm \tilde{c}_i$) configurations on this patch, so that the Chern-Simons integral of the entire modified Brillouin zone, denoted as $\tilde{T}^2$, is related to the Chern-Simons integral of a system with $c_i^\text{(tot)}=c_i + \tilde{c}_i = 0$. Since for the $c_i^\text{(tot)}=0$ system the Chern-Simons integral is gauge independent, we could use the Chern-Simons integral for this $c_i^\text{(tot)}=0$ system, whose domain of integration is a modified 3-torus $\tilde{T}^3 = \tilde{T}^2 \times S^1$, to define the linking invariant $\nu$, which equals exactly the Chern number of the post-quench Hamiltonian $c_f^\text{(tot)}=c_f + \tilde{c}_f$~\cite{Ce}. From this rather abstract argument, we get the linking invariant for the (2+1)-dimensional quench process as 
\begin{equation}\label{Eq:nu_cf-ci}
 	\nu  = c_f^{{\text{(tot)}}} = {c_f} + {{\tilde c}_f} = {c_f} \pm {{\tilde c}_i} = {c_f} \mp {c_i}.
 \end{equation} 
In the following, we perform concrete calculation of the Chern-Simons integral evaluated on the modified domain of integration $\tilde{T}^3 $ to revalidate Eq.~\eqref{Eq:nu_cf-ci}. In addition to that, as will be shown in the next section that the linking invariant $2nc_i$ ($n$ is an arbitrary integer) is always topologically trivial, we then arrive at our main results as shown in Eq.~\eqref{eq:mainrslt}. 

To simplify the problem, we choose a quasimomentum point $\mathbf{s}$ as the discontinuous point of the Berry connection, such that $\mathbf{h}(\mathbf{s})$ is either parallel or anti-parallel to $\mathbf{n}(\mathbf{s}, t = 0)$. For the case $c_i\neq c_f$, it is always possible to find such point; see, e.g., \cite{YiWei2019}. On the other hand, for the case $c_i=c_f\neq 0$, the Brouwer fixed point theorem (see, e.g., \cite{Hatcher2005algebraic}) shows that there exists at least one point such that either $\mathbf{n}(\mathbf{s}, t = 0)\cdot \mathbf{h}= 1$ or $\mathbf{n}(\mathbf{s}, t = 0)\cdot \mathbf{h}=- 1$. We then subtract an (infinitesimal) small disk around this point, which leaves a boundary, and cling a small $D^2$ (which is a 2-sphere subtracted by one point) with Skrymion spin configuration to that boundary so that $\mathbf{n}$ and $\mathbf{h}$ on the boundary of the small $D^2$ are continuous (see Fig.~\ref{Fig3} for an illustration of the clinging operation for the $|c_i|=1$ case). We can furthermore require that the magnetic field $\tilde{\mathbf{h}}$ on $D^2$ is always parallel or antiparallel to the corresponding spin configuration $\tilde{\mathbf{n}}$. By doing so, the preimage of a point on $S^2$
around $\bf{s}$ is always a straight line along the time direction. 
With respect to the original Brillouin zone $T^2$ (without considering the complementary patch), this {\lq\lq}clinging{\rq\rq} operation is equivalent to choosing a fixed gauge such that the Dirac string is either parallel or anti-parallel to the local magnetic field. We now proceed to perform concrete calculations. 

Assuming that the post-quench Hamiltonian is diagonalized as
\begin{align}
{\cal H}_f = {1\over 2} V\sigma_z V^\dagger,
\end{align}
the Berry connection then takes the following form:
\begin{align}
{\cal A}_\mu=-i\langle \xi_0|V e^{i\sigma_z t/2}V^\dagger \partial_\mu(V e^{-i\sigma_z t/2}V^\dagger|\xi_0\rangle).
\end{align}
Using this definition, the Berry connection could be calculated 
\begin{align}
\left\{\begin{aligned}
{\cal A}_t&=-{1\over 2}\langle \bar{\xi}|\sigma_z|\bar{\xi}\rangle={1\over 2}\bar{n}_z,\\
{\cal A}_i&=-i\langle \bar{\xi}| e^{i\sigma_z t/2}V^\dagger (\partial_i V) e^{-i\sigma_z t/2}|\bar{\xi}\rangle + \bar{\cal A}_i,
\end{aligned}
\right.
\end{align}
with $i=k_x, k_y$. 
Here, we denote $|\bar{\xi}\rangle=V^\dagger|\xi_0\rangle $ and $\bar{n}_i = \langle\bar{\xi}|\sigma_i|\bar{\xi}\rangle$. $\bar{\cal A}_\mu=-i\langle\bar{\xi}|\partial_\mu|\bar{\xi}\rangle$ and $\bar{\cal J}_{\mu}=\epsilon^{\mu\nu\rho}\partial_\nu{\bar{\cal A}_\rho}$ refer to the Berry connection and the Berry curvature, respectively, calculated from $|\bar{\xi}\rangle$. The discontinuity of $V$ is chosen to be the same as $|\xi_0\rangle$, namely $V(k_x, k_y)$ and $|\xi_0(k_x,k_y)\rangle$ are both discontinuous at $\mathbf{s} = (s_x,s_y)$, and $|\xi_0(s_x,s_y)\rangle$ is one of the eigenvector of $V(s_x,s_y)\sigma_z V^\dagger(s_x,s_y)$.

Then we see that ${\cal A}_i$ could be separated into two parts: One is time-independent, and the other is time-dependent.
We write $V=\left(|1\rangle\;\;|2\rangle\right)$, and the gauge could be chosen so that $\langle1|\partial_i|1\rangle=-\langle2|\partial_i|2\rangle$. 
Then the time independent part of the Berry connection is
\begin{align}
{\cal A}_i^0=-i\langle1|\partial_i|1\rangle \langle\bar{\xi}|\sigma_z|\bar{\xi}\rangle + \bar{\cal A}_i,
\end{align}
and the time-dependent part is
\begin{align}
{\cal A}_i^t=-i\langle1|\partial_i|2\rangle \exp(-it) \langle\bar{\xi}|\sigma_+|\bar{\xi}\rangle + \text{c.c.},
\end{align}
where $\text{c.c.}$ means complex conjugate. 

Using these relations, we could calculate the Chern-Simons integral. In the following part, we calculate the Chern-Simons integral in the region outside the small $D^2$, then the domain of integration has a boundary. We denote the domain of integration for the quasimomentum as $M=T^2\setminus \bf{s}$,
 and the Chern Simons integral is evaluated on the manifold $N=M\times S^1$. The result is 
\begin{align}
&\int_N d^3x {\cal A}_t{\cal J}_{t}={\pi}\int_M d^2x  {\cal J}^f_{t} \bar{n}_z^2+{\pi} \oint_{\partial M} \bar {n}_z^2{\cal A}^f\cdot dx,\\
&\begin{aligned}
\int_N d^3x \epsilon^{ij}{\cal A}_i\partial_j {\cal A}_t =&{\pi}  \int_M d^2x {\cal J}_{t}^f \bar{n}^2_z-{\pi} \oint_{\partial M} \bar {n}_z^2{\cal A}^f\cdot dx  \\ 
&-2\pi\oint_{\partial M} \bar{n}_z\bar{\cal A}\cdot dx,
\end{aligned}\\
&\int_N d^3x \epsilon^{ij}{\cal A}_i\partial_t{\cal A}_j = -2\pi\int_M d^2x {\cal J}^f_{t} (\bar{n}^2_x + \bar{n}^2_y).
\end{align}
Here, $\epsilon^{ij}$ is the 2-dimensional Levi-Civita symbol. ${\cal A}^f$ and ${\cal J}^f $ are defined, respectively, as the Berry connection and Berry curvature of the post-quench Hamiltonian, i.e., ${\cal A}^f_i=-i\langle1|\partial_i|1\rangle$ and ${\cal J}_{t}^f=\partial_x {\cal A}^f_y-\partial_y{\cal A}^f_x$. 

The Chern Simons integral on the manifold $N$ is
\begin{align}
\frac{1}{4\pi^2}\int_N \epsilon^{\mu\nu\rho}{\cal A}_\mu\partial_\nu {\cal A}_\rho= -\frac{c_i}{2}\bar{n}_z^{\text{Dirac}} + c_f. \label{eq:FR} 
\end{align}
Here, $\bar{n}_z^{\text{Dirac}}$ comes from the integral $\oint_{\partial M} \bar{n}_z \bar{\cal A}\cdot dx$. Since the boundary $\partial_M$ is an infinitesimal circle surrounding the Dirac string direction $\mathbf{s}$, $\bar{n}_z=\bar{n}_z(\mathbf{s}) $. In particular, we have $\bar{n}_z^\text{Dirac}=\pm 1$ due to the fact that the position $\mathbf{s}$ is chosen as the point such that $\mathbf{n}(\mathbf{s}, t=0)$ is parallel or antiparallel to $\mathbf{h}(\mathbf{s})$. When considering the constant-magnetic-field problem (with $c_i=1$, $c_f=0$) as described in the last section, we find ${1\over 4\pi^2}\int_N \epsilon^{\mu\nu\rho}{\cal A}_\mu\partial_\nu {\cal A}_\rho = -{1\over 2} c_i \bar{n}_z^\text{Dirac}$, which is consistent with Eq.~\eqref{eq:constH} with $c_i = 1$, $\bar{n}_z^\text{Dirac} = \pm1$ (the Dirac string live at the North or South Poles of $S^2$) when $\mathbf{d} = \pm \mathbf{h}$.

According to \eqref{eq:FR}, by changing the domain of integration from $N=({T^2}\backslash {\mathbf{s}})\times S^1$ to ${D^2} \times {S^1}$ which share the same boundary, we get the Chern-Simons integral from the complementary part 
(by substituting $c_i\to \tilde{c}_i$ and $c_f\to \tilde{c}_f$) as follows 
\begin{equation}
	\frac{1}{{4{\pi ^2}}}\int_{{D^2} \times {S^1}} {{\epsilon ^{\mu \nu \rho }}} {\mathcal{A}_\mu }{\partial _\nu }{\mathcal{A}_\rho } =  - \frac{{\tilde c}_i}{2}\bar n_z^{{\text{Dirac}}} + {{\tilde c}_f} =  - \frac{c_i}{2}\bar n_z^{{\text{Dirac}}},
\end{equation}
where we set $\tilde{c}_i=-c_i$ in order to compensate the Chern number and $\tilde{c}_f=\tilde{c}_i\bar{n}_z^{\text{Dirac}}$ to make the spin configuration and the evolving Hamiltonian continuous at the clinging point $\mathbf{s}$. Then the Chern-Simons integral for the whole region $\tilde{T}^3 = \tilde{T}^2 \times S^1 $ [where ${{\tilde T}^2} = ({T^2}\backslash {\mathbf{s}}) \cup {D^2}$ is the modified Brillouin zone] is
\begin{align} \label{Eq:I_tilde_CS}
\tilde{I}_{CS}=\frac{1}{4\pi^2}\int_{\tilde{T}^3} \epsilon^{\mu\nu\rho}{\cal A}_\mu\partial_\nu {\cal A}_\rho =- c_i\bar{n}_z^{\text{Dirac}}+c_f.
\end{align}
The Chern-Simons integral on $\tilde{T}^3$ can be identified as the linking invariant $\nu$ for the (2+1)-dimensional quench process:
\begin{equation} \label{Eq:nu_I_CS}
	\nu  = {{\tilde I}_{{\text{CS}}}} = {c_f} \mp {c_i},
\end{equation}
where $\bar{n}_z^{\text{Dirac}} = \pm 1$ has been used in Eq.~\eqref{Eq:I_tilde_CS}. We thus recover Eq.~\eqref{Eq:nu_cf-ci} as expected. 

\subsection{Triviality of linking invariant $2nc_i$}

\begin{figure}[bth]
\includegraphics[width=\columnwidth]{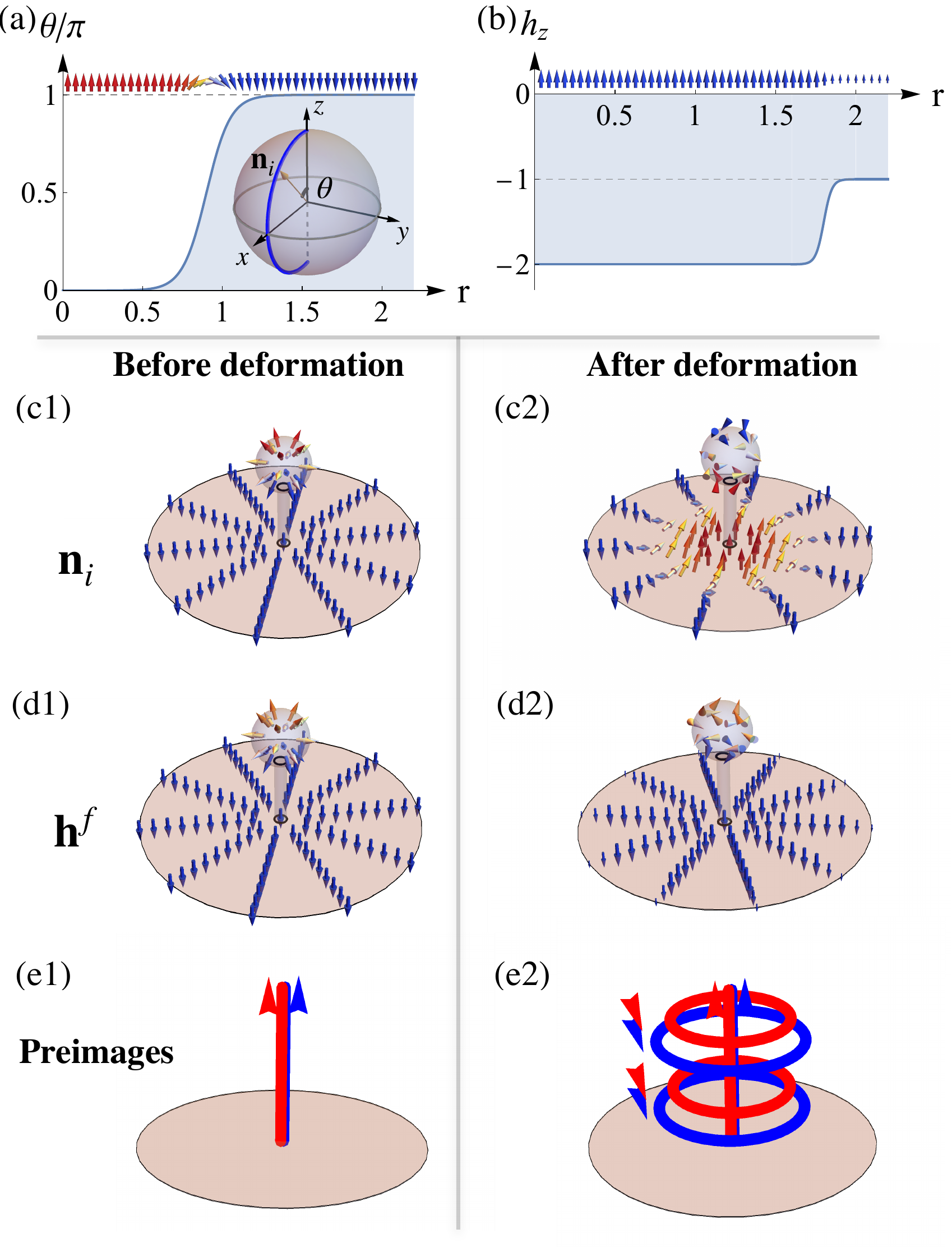}
\caption{\label{Fig4} 
Continuous deformation of the spin configuration $\mathbf{n}_i$ (a) and magnetic field $\mathbf{h}^f$ (b) on the disk. Before the deformation, both $\mathbf{n}_i$ (c1) and $\mathbf{h}^f$ (d1) are uniform on the disk, pointing to the South Pole direction. After the deformation, $\mathbf{n}_i(r, \Phi)$ is rotated to the direction $\mathbf{n}_i(r, \Phi) = (\sin\theta(r), 0, \cos\theta(r))$ [with $\theta(r)$ given by (a)] (c2), and the magnitude of $\mathbf{h}^f$ is increased as $|\mathbf{h}^f(r, \phi)| = h_z(r)$ [with $h_z(r)$ given by (b)] (d2). The spin and magnetic field configurations for the attached $D^2$ in (c2) and (d2) are also deformed to make them continuous at the clinging point: Compared with (c1) and (d1), the corresponding vector components are deformed as $x\to x, y\to-y, z\to -z$ on $D^2$. We see that, the continuous deformations of $\mathbf{n}_i$ and $\mathbf{h}^f$ give rise to a change of linking numbers (for any two preimages of any two points, except for the North or South poles, on the Bloch sphere $S^2$) from $0$ (e1) to $2nc_i$ (which equals $4$ for the current case with $n=2$ and $c_i=1$) (e2).
}
\end{figure}

In the last section, we have shown that the linking number of $\nu=c_f+c_i$ is equivalent to $c_f-c_i$. In this section, we will construct a continuous mapping to show that the linking number of $2nc_i$ is equivalent to the trivial case with no linking, and, as a consequence, the linking invariant is only defined module $2c_i$ as shown in Eq.~\eqref{eq:mainrslt}. 

Since there always exists a wavevector $\mathbf{s}$ such that ${\bf h}({\bf s})={\bf n}({\bf s}, t = 0)$ or ${\bf h}({\bf s})=-{\bf n}({\bf s})$ as mentioned in the last section, we continuously deform the spin configuration and magnetic field configuration around $\bf s$ so that the new quench process is topologically equivalent to the initial one. The deformation includes three steps. We suppose as well that ${\bf n}({\bf s}, t = 0)={\bf h}({\bf s})$ points to the South Pole direction of the Bloch sphere. 

\begin{itemize}
	\item Firstly, we subtract an infinitesimal disk around $\bf s$, which leaves an edge in the rest manifold, and cling to the edge a finite disk with spin configuration polarized in ${\bf n}({\bf s})$ direction [Fig.~\ref{Fig4}(c1)] and magnetic field polarized in ${\bf h}({\bf s})$ direction [Fig.~\ref{Fig4}(d1)]. The purpose of the first step is to expand the point ${\bf s}$ to a disk region.
    \item Secondly, we deform the spin configuration in the interior of the disk [Figs.~\ref{Fig4}(a) and \ref{Fig4}(c2)]. In this step we assume that the spin configuration only depends on the distance $r$ from the center of the disk: $\mathbf{n}_i(r, \Phi) = (\sin\theta(r), 0, \cos\theta(r))$ [see inset of Fig.~\ref{Fig4}(a)], where $r$ and $\Phi$ are the polar coordinates for the disk. The typical form of $\theta(r)$ is shown in Fig.~\ref{Fig4}(a), and the corresponding spin configuration for a particular $\Phi$ is indicated by the arrows above the $\theta(r)$ curve. This assumptions implies that the total Berry curvature over the disk is zero, because for every point on the circle with the same $r$, 
    the spins point to the same direction as shown in Fig.~\ref{Fig4}(c2). We require the magnetic field on the disk is along the $z$ direction and then every point on the Bloch sphere has preimages in this disk. Assuming that the magnetic field on the disk is rotationally symmetric with respect to the center of the disk, the preimages of each point on the Bloch sphere are certain circles parallel to the $(k_x,k_y)$ plane.  
	\item Finally, we deform the magnitude of the magnetic field and break the requirement of $|{\bf h}^f|=1$ at this stage [Figs.~\ref{Fig4}(b) and \ref{Fig4}(d2)]. In the second step, if the magnetic field is constant, namely $|{\bf h}^f|=1$, the preimage loops for each point on the Bloch sphere only appear once. In order to obtain more than one preimage loop, we propose to deform the magnitude of the magnetic field. The deformation of the magnitude of the magnetic field is shown in Fig. \ref{Fig4}(b). The requirement of $|{\bf h}^f|=1$ is proposed to make the quench process periodic in the time direction so that the manifold of $(k_x,k_y,t)$ is compact. However, when ${\bf h}^f$ is parallel or anti-parallel to ${\bf n}_i$, the spin ${\bf n}(k_x, k_y, t) $ stays unchanged and is periodic at $t=2\pi$ under \emph{any} strength of magnetic field. In the interior of the disk, the strength of the magnetic field keeps increasing until the norm of the magnetic field is an integer $n$ multiple of the norm of magnetic field outside this disk, so that the magnetic field of the interior of the disk makes the evolving of spins with arbitrary direction periodic at $t=2\pi$. 
	The resulting magnetic field configuration with $n=2$ after such deformation is shown in Fig.~\ref{Fig4}(d2). 
\end{itemize}

These three steps do not change the topological properties of the quench process, since all the three steps are operated continuously.  
As shown in Figs.~\ref{Fig4} (e1) and (e2), we see that the continuous deformation introduces a change of linking number $2 n c_i$ (with $c_i = 1$, $n = 2$ for the current case) for the preimages of two points on $S^2$. Thus we see that the linking invariant $2n c_i$ is always topologically trivial for the quench process with initial Chern number $c_i$. Combining this observation together with the results of Eq.~\eqref{Eq:nu_I_CS}, we have then proved that the topological invariant in the (2+1)-dimensional quench process is the linking invariant $\nu = (c_f-c_i)\mod (2c_i)$ as we have presented in Eq.~\eqref{eq:mainrslt}. 

\section{Experimental feasibility}\label{Sec:Experimental-feasibility}
We note that the motivation of our previous studies on the quench dynamics starting with a trivial initial state~\cite{Ce,Yu2017} can be partially attributed to the fact that a topological initial state for the ultracold quantum gases could not be prepared reliably by adiabatically tuning the controlling parameter(s), e.g., the mass term $M$ of the quantum anomalous Hall model in Eq.~\eqref{Eq:Hami_QAH} above, because as we tune it to a critical value, the spectral gap closes and adiabaticity breaks down. This problem is resolved in a very recent work~\cite{Simone-Jinlong-Peter-Jan2019}, which shows that topological Chern insulator states can also be prepared adiabatically for ultracold quantum gases by introducing a symmetry-conjugated duplicate of the target system. As elaborated in~\cite{Simone-Jinlong-Peter-Jan2019}, starting from a trivial product state $\lvert c=0 \rangle_{\rm S}\otimes\lvert  c=0\rangle_\text{S*}$ for both the system (S) and its conjugated duplicate (S*) with trivial Chern numbers, by intermittently introducing symmetry-breaking couplings, it is possible to find a \emph{gapped} path connecting this trivial product state $\lvert c=0 \rangle_{\rm S}\otimes\lvert  c=0\rangle_\text{S*}$ to a nontrivial product state $\lvert c=1 \rangle_{\rm S}\otimes\lvert  c=-1\rangle_\text{S*}$. One can then project the decoupled total system onto the target system S by selectively removing S* using, e.g., a magnetic field gradient~\cite{TopHaldane}, leaving alone the topological state $\lvert c=1 \rangle_{\rm S}$ for the target system to perform further experimental investigations, e.g., to perform quench dynamics as discussed in the current work. This indicates that our current protocol to probe the linking invariant of the torus homotopy group for the quench dynamics in ultracold quantum gases is ready to be implemented with state-of-the-art experiments.

\section {Conclusion} \label{Sec:Conclusion}
In conclusion, we have studied the topological invariant in the quench dynamics of a 2D two-band Chern insulator with arbitrary initial Chern number $c_i$ and post-quench Chern number $c_f$. The topological invariant is the linking invariant $\nu$, which is determined by the difference between these two Chern numbers module $2c_i$: $\nu = (c_f-c_i)\mod (2c_i)$. We have made use of concrete examples to validate this results, and then performed concrete calculation of the Chern-Simons integral defined on a modified quasimomentum-time manifold to evaluate the linking invariant. Provided a suitable preparation of a topological initial state~\cite{Simone-Jinlong-Peter-Jan2019}, 
together with suitable Bloch-state tomography~\cite{Yu2019Hopf,Yi-Shuai2019}, the linking structure of preimage loops as well as the linking invariant from the modified Chern-Simons integral can be experimentally measured. 
They can also be directly measured by certain solid-state simulators, e.g., the nitrogen-vacancy centers, as described recently~\cite{Duan2017}.  

\begin{acknowledgements}
We would like to thank Hui Zhai, Pengfei Zhang, and Ning Sun for enlightening discussions. This work is supported by MOST under Grant No. 2016YFA0301600 and NSFC Grant No. 11734010. 
\end{acknowledgements}

\appendix
\section{Chern number of the $(t,k_x)$ torus} \label{App.A}
In this section, we will show that the Chern numbers of the $(t,k_x)$ torus and $(t,k_y)$ torus are zero. We take the $(k_x, t)$ torus as an example. 

Due to the fact that the post-quench Hamiltonian is time-independent, the spin direction has the following relation,
\begin{align}
\partial_t {\bf n} &= {\bf h}^f\times {\bf n},\\
\frac{1}{T}\int_0^T dt {\bf n}&=  ({\bf h}^i\cdot {\bf h}^f){\bf h}^f,
\end{align}
where $T=2\pi$ is the evolving period. 

The Berry curvature of the $(k_x, t)$ torus is calculated as follows:
\begin{align}
{\cal J}_{y} = {1\over 2}{\bf n}\cdot(\partial_x{\bf n}\times\partial_t{\bf n}) = -\partial_x {\bf n}\cdot {\bf h}^f.\label{eq:F0x}
\end{align}
Then we can show that the corresponding Chern number $c_y$ vanishes:
\begin{align}
c_y = -\frac{1}{4\pi}\int_{S^1} dx {\bf h}^f\cdot \int_{S^1} dt \partial_x {\bf n} = 0.\label{eq:C0x}
\end{align}
In  Eqs. \eqref{eq:F0x} and \eqref{eq:C0x}, we have made use of the fact that $\partial_x\mathbf{h}^f\cdot\mathbf{h}^{f}=0$ and $\mathbf{n}\cdot \partial_x\mathbf{n}=0$, due to the normalization conditions $|\mathbf{h}^f|^2 = 1$ and $|\mathbf{n}|^2=1$.

In the same way, we can also show that $c_x = 0$. 

\section{Chern-Simons integral of constant magnetic field\label{sec:CSGauge}}
In this section, we will show that the Chern-Simons integral for a Skyrmion in a constant magnetic field~\cite{Azee} is gauge dependent.

Since a uniform magnetic field is independent of the quasimomentum $(k_x, k_y)$, it could be proved that the Berry curvature is independent of $t$. 

Firstly, we see that
\begin{align}
\int_{T^3} d^3x{\cal J}_{t}{\cal A}_t = -\pi{\bf h}^f\cdot\int d^2x{\cal J}_{t} {\bf n} = 0.
\end{align}
This is mainly due to the fact that in a constant magnetic field ${\cal J}_{t}$ is independent of time $t$, and ${\cal J}_{t}={1\over 4}\mathbf{n}\cdot(\partial_x\mathbf{n}\times\partial_y\mathbf{n}) $.

Since the Berry connection is irrelevant of time $t$, we could choose a gauge so that the Berry connection is also irrelevant of time: $\partial_t A_i = 0$ with $ i={x,y}$. 
We denote the direction of the Dirac string, which is the singularity of the Berry connection, as ${\bf d}$. Due to the singularity of Berry connection, the integral is now defined on a manifold with boundary:
\begin{align}
4\pi^2I_\text{CS}=\int_{T^3\setminus {\bf d}} d^3x \epsilon^{ij}{\cal A}_i\partial_j{\cal A}_t =-2\pi^2 {\bf d}\cdot {\bf h}^f,\label{eq:constHICS}
\end{align}
which depends on the gauge chosen. The right hand side of Eq. \eqref{eq:constHICS} originates from the boundary term of the integration by parts.


\begin{thebibliography}{42}%
\makeatletter
\providecommand \@ifxundefined [1]{%
 \@ifx{#1\undefined}
}%
\providecommand \@ifnum [1]{%
 \ifnum #1\expandafter \@firstoftwo
 \else \expandafter \@secondoftwo
 \fi
}%
\providecommand \@ifx [1]{%
 \ifx #1\expandafter \@firstoftwo
 \else \expandafter \@secondoftwo
 \fi
}%
\providecommand \natexlab [1]{#1}%
\providecommand \enquote  [1]{``#1''}%
\providecommand \bibnamefont  [1]{#1}%
\providecommand \bibfnamefont [1]{#1}%
\providecommand \citenamefont [1]{#1}%
\providecommand \href@noop [0]{\@secondoftwo}%
\providecommand \href [0]{\begingroup \@sanitize@url \@href}%
\providecommand \@href[1]{\@@startlink{#1}\@@href}%
\providecommand \@@href[1]{\endgroup#1\@@endlink}%
\providecommand \@sanitize@url [0]{\catcode `\\12\catcode `\$12\catcode
  `\&12\catcode `\#12\catcode `\^12\catcode `\_12\catcode `\%12\relax}%
\providecommand \@@startlink[1]{}%
\providecommand \@@endlink[0]{}%
\providecommand \url  [0]{\begingroup\@sanitize@url \@url }%
\providecommand \@url [1]{\endgroup\@href {#1}{\urlprefix }}%
\providecommand \urlprefix  [0]{URL }%
\providecommand \Eprint [0]{\href }%
\providecommand \doibase [0]{http://dx.doi.org/}%
\providecommand \selectlanguage [0]{\@gobble}%
\providecommand \bibinfo  [0]{\@secondoftwo}%
\providecommand \bibfield  [0]{\@secondoftwo}%
\providecommand \translation [1]{[#1]}%
\providecommand \BibitemOpen [0]{}%
\providecommand \bibitemStop [0]{}%
\providecommand \bibitemNoStop [0]{.\EOS\space}%
\providecommand \EOS [0]{\spacefactor3000\relax}%
\providecommand \BibitemShut  [1]{\csname bibitem#1\endcsname}%
\let\auto@bib@innerbib\@empty
\bibitem [{\citenamefont {Haldane}(1988)}]{Haldane1988}%
  \BibitemOpen
  \bibfield  {author} {\bibinfo {author} {\bibfnamefont {F.~D.~M.}\
  \bibnamefont {Haldane}},\ }\bibfield  {title} {\enquote {\bibinfo {title}
  {{Model for a Quantum Hall Effect without Landau Levels: Condensed-Matter
  Realization of the 'Parity Anomaly'}},}\ }\href {\doibase
  10.1103/PhysRevLett.61.2015} {\bibfield  {journal} {\bibinfo  {journal}
  {Phys. Rev. Lett.}\ }\textbf {\bibinfo {volume} {61}},\ \bibinfo {pages}
  {2015} (\bibinfo {year} {1988})}\BibitemShut {NoStop}%
\bibitem [{\citenamefont {Thouless}\ \emph {et~al.}(1982)\citenamefont
  {Thouless}, \citenamefont {Kohmoto}, \citenamefont {Nightingale},\ and\
  \citenamefont {den Nijs}}]{TKNN1982}%
  \BibitemOpen
  \bibfield  {author} {\bibinfo {author} {\bibfnamefont {D.~J.}\ \bibnamefont
  {Thouless}}, \bibinfo {author} {\bibfnamefont {M.}~\bibnamefont {Kohmoto}},
  \bibinfo {author} {\bibfnamefont {M.~P.}\ \bibnamefont {Nightingale}}, \ and\
  \bibinfo {author} {\bibfnamefont {M.}~\bibnamefont {den Nijs}},\ }\bibfield
  {title} {\enquote {\bibinfo {title} {{Quantized Hall Conductance in a
  Two-Dimensional Periodic Potential}},}\ }\href {\doibase
  10.1103/PhysRevLett.49.405} {\bibfield  {journal} {\bibinfo  {journal} {Phys.
  Rev. Lett.}\ }\textbf {\bibinfo {volume} {49}},\ \bibinfo {pages} {405}
  (\bibinfo {year} {1982})}\BibitemShut {NoStop}%
\bibitem [{\citenamefont {Qi}\ \emph {et~al.}(2006)\citenamefont {Qi},
  \citenamefont {Wu},\ and\ \citenamefont {Zhang}}]{Qi2006}%
  \BibitemOpen
  \bibfield  {author} {\bibinfo {author} {\bibfnamefont {X.-L.}\ \bibnamefont
  {Qi}}, \bibinfo {author} {\bibfnamefont {Y.-S.}\ \bibnamefont {Wu}}, \ and\
  \bibinfo {author} {\bibfnamefont {S.-C.}\ \bibnamefont {Zhang}},\ }\bibfield
  {title} {\enquote {\bibinfo {title} {{Topological quantization of the spin
  Hall effect in two-dimensional paramagnetic semiconductors}},}\ }\href
  {\doibase 10.1103/PhysRevB.74.085308} {\bibfield  {journal} {\bibinfo
  {journal} {Phys. Rev. B}\ }\textbf {\bibinfo {volume} {74}},\ \bibinfo
  {pages} {085308} (\bibinfo {year} {2006})}\BibitemShut {NoStop}%
\bibitem [{\citenamefont {Jotzu}\ \emph {et~al.}(2014)\citenamefont {Jotzu},
  \citenamefont {Messer}, \citenamefont {Desbuquois}, \citenamefont {Lebrat},
  \citenamefont {Uehlinger}, \citenamefont {Greif},\ and\ \citenamefont
  {Esslinger}}]{TopHaldane}%
  \BibitemOpen
  \bibfield  {author} {\bibinfo {author} {\bibfnamefont {G.}~\bibnamefont
  {Jotzu}}, \bibinfo {author} {\bibfnamefont {M.}~\bibnamefont {Messer}},
  \bibinfo {author} {\bibfnamefont {R.}~\bibnamefont {Desbuquois}}, \bibinfo
  {author} {\bibfnamefont {M.}~\bibnamefont {Lebrat}}, \bibinfo {author}
  {\bibfnamefont {T.}~\bibnamefont {Uehlinger}}, \bibinfo {author}
  {\bibfnamefont {D.}~\bibnamefont {Greif}}, \ and\ \bibinfo {author}
  {\bibfnamefont {T.}~\bibnamefont {Esslinger}},\ }\bibfield  {title} {\enquote
  {\bibinfo {title} {{Experimental realization of the topological Haldane model
  with ultracold fermions}},}\ }\href {\doibase 10.1038/nature13915} {\bibfield
   {journal} {\bibinfo  {journal} {Nature (London)}\ }\textbf {\bibinfo
  {volume} {515}},\ \bibinfo {pages} {237} (\bibinfo {year}
  {2014})}\BibitemShut {NoStop}%
\bibitem [{\citenamefont {Wu}\ \emph {et~al.}(2016)\citenamefont {Wu},
  \citenamefont {Zhang}, \citenamefont {Sun}, \citenamefont {Xu}, \citenamefont
  {Wang}, \citenamefont {Ji}, \citenamefont {Deng}, \citenamefont {Chen},
  \citenamefont {Liu},\ and\ \citenamefont {Pan}}]{Wu2016}%
  \BibitemOpen
  \bibfield  {author} {\bibinfo {author} {\bibfnamefont {Z.}~\bibnamefont
  {Wu}}, \bibinfo {author} {\bibfnamefont {L.}~\bibnamefont {Zhang}}, \bibinfo
  {author} {\bibfnamefont {W.}~\bibnamefont {Sun}}, \bibinfo {author}
  {\bibfnamefont {X.-T.}\ \bibnamefont {Xu}}, \bibinfo {author} {\bibfnamefont
  {B.-Z.}\ \bibnamefont {Wang}}, \bibinfo {author} {\bibfnamefont {S.-C.}\
  \bibnamefont {Ji}}, \bibinfo {author} {\bibfnamefont {Y.}~\bibnamefont
  {Deng}}, \bibinfo {author} {\bibfnamefont {S.}~\bibnamefont {Chen}}, \bibinfo
  {author} {\bibfnamefont {X.-J.}\ \bibnamefont {Liu}}, \ and\ \bibinfo
  {author} {\bibfnamefont {J.-W.}\ \bibnamefont {Pan}},\ }\bibfield  {title}
  {\enquote {\bibinfo {title} {{Realization of two-dimensional spin-orbit
  coupling for Bose-Einstein condensates}},}\ }\href {\doibase
  10.1126/science.aaf6689} {\bibfield  {journal} {\bibinfo  {journal}
  {Science}\ }\textbf {\bibinfo {volume} {354}},\ \bibinfo {pages} {83}
  (\bibinfo {year} {2016})}\BibitemShut {NoStop}%
\bibitem [{\citenamefont {Caio}\ \emph {et~al.}(2015)\citenamefont {Caio},
  \citenamefont {Cooper},\ and\ \citenamefont {Bhaseen}}]{Caio2015}%
  \BibitemOpen
  \bibfield  {author} {\bibinfo {author} {\bibfnamefont {M.~D.}\ \bibnamefont
  {Caio}}, \bibinfo {author} {\bibfnamefont {N.~R.}\ \bibnamefont {Cooper}}, \
  and\ \bibinfo {author} {\bibfnamefont {M.~J.}\ \bibnamefont {Bhaseen}},\
  }\bibfield  {title} {\enquote {\bibinfo {title} {{Quantum Quenches in Chern
  Insulators}},}\ }\href {\doibase 10.1103/PhysRevLett.115.236403} {\bibfield
  {journal} {\bibinfo  {journal} {Phys. Rev. Lett.}\ }\textbf {\bibinfo
  {volume} {115}},\ \bibinfo {pages} {236403} (\bibinfo {year}
  {2015})}\BibitemShut {NoStop}%
\bibitem [{\citenamefont {Caio}\ \emph {et~al.}(2016)\citenamefont {Caio},
  \citenamefont {Cooper},\ and\ \citenamefont {Bhaseen}}]{Caio2016}%
  \BibitemOpen
  \bibfield  {author} {\bibinfo {author} {\bibfnamefont {M.~D.}\ \bibnamefont
  {Caio}}, \bibinfo {author} {\bibfnamefont {N.~R.}\ \bibnamefont {Cooper}}, \
  and\ \bibinfo {author} {\bibfnamefont {M.~J.}\ \bibnamefont {Bhaseen}},\
  }\bibfield  {title} {\enquote {\bibinfo {title} {{Hall response and edge
  current dynamics in Chern insulators out of equilibrium}},}\ }\href {\doibase
  10.1103/PhysRevB.94.155104} {\bibfield  {journal} {\bibinfo  {journal} {Phys.
  Rev. B}\ }\textbf {\bibinfo {volume} {94}},\ \bibinfo {pages} {155104}
  (\bibinfo {year} {2016})}\BibitemShut {NoStop}%
\bibitem [{\citenamefont {Hu}\ \emph {et~al.}(2016)\citenamefont {Hu},
  \citenamefont {Zoller},\ and\ \citenamefont {Budich}}]{Hu-Zoller2016}%
  \BibitemOpen
  \bibfield  {author} {\bibinfo {author} {\bibfnamefont {Y.}~\bibnamefont
  {Hu}}, \bibinfo {author} {\bibfnamefont {P.}~\bibnamefont {Zoller}}, \ and\
  \bibinfo {author} {\bibfnamefont {J.~C.}\ \bibnamefont {Budich}},\ }\bibfield
   {title} {\enquote {\bibinfo {title} {{Dynamical Buildup of a Quantized Hall
  Response from Nontopological States}},}\ }\href {\doibase
  10.1103/PhysRevLett.117.126803} {\bibfield  {journal} {\bibinfo  {journal}
  {Phys. Rev. Lett.}\ }\textbf {\bibinfo {volume} {117}},\ \bibinfo {pages}
  {126803} (\bibinfo {year} {2016})}\BibitemShut {NoStop}%
\bibitem [{\citenamefont {\"Unal}\ \emph {et~al.}(2016)\citenamefont {\"Unal},
  \citenamefont {Mueller},\ and\ \citenamefont {Oktel}}]{Unal2016}%
  \BibitemOpen
  \bibfield  {author} {\bibinfo {author} {\bibfnamefont {F.~N.}\ \bibnamefont
  {\"Unal}}, \bibinfo {author} {\bibfnamefont {E.~J.}\ \bibnamefont {Mueller}},
  \ and\ \bibinfo {author} {\bibfnamefont {M.~O.}\ \bibnamefont {Oktel}},\
  }\bibfield  {title} {\enquote {\bibinfo {title} {{Nonequilibrium fractional
  Hall response after a topological quench}},}\ }\href {\doibase
  10.1103/PhysRevA.94.053604} {\bibfield  {journal} {\bibinfo  {journal} {Phys.
  Rev. A}\ }\textbf {\bibinfo {volume} {94}},\ \bibinfo {pages} {053604}
  (\bibinfo {year} {2016})}\BibitemShut {NoStop}%
\bibitem [{\citenamefont {Wilson}\ \emph {et~al.}(2016)\citenamefont {Wilson},
  \citenamefont {Song},\ and\ \citenamefont {Refael}}]{Wilson2016}%
  \BibitemOpen
  \bibfield  {author} {\bibinfo {author} {\bibfnamefont {J.~H.}\ \bibnamefont
  {Wilson}}, \bibinfo {author} {\bibfnamefont {J.~C.~W.}\ \bibnamefont {Song}},
  \ and\ \bibinfo {author} {\bibfnamefont {G.}~\bibnamefont {Refael}},\
  }\bibfield  {title} {\enquote {\bibinfo {title} {{Remnant Geometric Hall
  Response in a Quantum Quench}},}\ }\href {\doibase
  10.1103/PhysRevLett.117.235302} {\bibfield  {journal} {\bibinfo  {journal}
  {Phys. Rev. Lett.}\ }\textbf {\bibinfo {volume} {117}},\ \bibinfo {pages}
  {235302} (\bibinfo {year} {2016})}\BibitemShut {NoStop}%
\bibitem [{\citenamefont {Wang}\ \emph {et~al.}(2017)\citenamefont {Wang},
  \citenamefont {Zhang}, \citenamefont {Chen}, \citenamefont {Yu},\ and\
  \citenamefont {Zhai}}]{Ce}%
  \BibitemOpen
  \bibfield  {author} {\bibinfo {author} {\bibfnamefont {C.}~\bibnamefont
  {Wang}}, \bibinfo {author} {\bibfnamefont {P.}~\bibnamefont {Zhang}},
  \bibinfo {author} {\bibfnamefont {X.}~\bibnamefont {Chen}}, \bibinfo {author}
  {\bibfnamefont {J.}~\bibnamefont {Yu}}, \ and\ \bibinfo {author}
  {\bibfnamefont {H.}~\bibnamefont {Zhai}},\ }\bibfield  {title} {\enquote
  {\bibinfo {title} {{Scheme to Measure the Topological Number of a Chern
  Insulator from Quench Dynamics}},}\ }\href {\doibase
  10.1103/PhysRevLett.118.185701} {\bibfield  {journal} {\bibinfo  {journal}
  {Phys. Rev. Lett.}\ }\textbf {\bibinfo {volume} {118}},\ \bibinfo {pages}
  {185701} (\bibinfo {year} {2017})}\BibitemShut {NoStop}%
\bibitem [{\citenamefont {Yu}(2017)}]{Yu2017}%
  \BibitemOpen
  \bibfield  {author} {\bibinfo {author} {\bibfnamefont {J.}~\bibnamefont
  {Yu}},\ }\bibfield  {title} {\enquote {\bibinfo {title} {{Phase vortices of
  the quenched Haldane model}},}\ }\href {\doibase 10.1103/PhysRevA.96.023601}
  {\bibfield  {journal} {\bibinfo  {journal} {Phys. Rev. A}\ }\textbf {\bibinfo
  {volume} {96}},\ \bibinfo {pages} {023601} (\bibinfo {year}
  {2017})}\BibitemShut {NoStop}%
\bibitem [{\citenamefont {Chang}(2018)}]{Chang2018}%
  \BibitemOpen
  \bibfield  {author} {\bibinfo {author} {\bibfnamefont {P.-Y.}\ \bibnamefont
  {Chang}},\ }\bibfield  {title} {\enquote {\bibinfo {title} {Topology and
  entanglement in quench dynamics},}\ }\href {\doibase
  10.1103/PhysRevB.97.224304} {\bibfield  {journal} {\bibinfo  {journal} {Phys.
  Rev. B}\ }\textbf {\bibinfo {volume} {97}},\ \bibinfo {pages} {224304}
  (\bibinfo {year} {2018})}\BibitemShut {NoStop}%
\bibitem [{\citenamefont {Ezawa}(2018)}]{Ezawa}%
  \BibitemOpen
  \bibfield  {author} {\bibinfo {author} {\bibfnamefont {M.}~\bibnamefont
  {Ezawa}},\ }\bibfield  {title} {\enquote {\bibinfo {title} {{Topological
  quantum quench dynamics carrying arbitrary Hopf and second Chern numbers}},}\
  }\href {\doibase 10.1103/PhysRevB.98.205406} {\bibfield  {journal} {\bibinfo
  {journal} {Phys. Rev. B}\ }\textbf {\bibinfo {volume} {98}},\ \bibinfo
  {pages} {205406} (\bibinfo {year} {2018})}\BibitemShut {NoStop}%
\bibitem [{\citenamefont {Qiu}\ \emph {et~al.}(2018)\citenamefont {Qiu},
  \citenamefont {Deng}, \citenamefont {Guo},\ and\ \citenamefont
  {Yi}}]{Yi2018A}%
  \BibitemOpen
  \bibfield  {author} {\bibinfo {author} {\bibfnamefont {X.}~\bibnamefont
  {Qiu}}, \bibinfo {author} {\bibfnamefont {T.-S.}\ \bibnamefont {Deng}},
  \bibinfo {author} {\bibfnamefont {G.-C.}\ \bibnamefont {Guo}}, \ and\
  \bibinfo {author} {\bibfnamefont {W.}~\bibnamefont {Yi}},\ }\bibfield
  {title} {\enquote {\bibinfo {title} {Dynamical topological invariants and
  reduced rate functions for dynamical quantum phase transitions in two
  dimensions},}\ }\href {\doibase 10.1103/PhysRevA.98.021601} {\bibfield
  {journal} {\bibinfo  {journal} {Phys. Rev. A}\ }\textbf {\bibinfo {volume}
  {98}},\ \bibinfo {pages} {021601(R)} (\bibinfo {year} {2018})}\BibitemShut
  {NoStop}%
\bibitem [{\citenamefont {Zhang}\ \emph {et~al.}(2018)\citenamefont {Zhang},
  \citenamefont {Zhang}, \citenamefont {Niu},\ and\ \citenamefont
  {Liu}}]{Liu2018}%
  \BibitemOpen
  \bibfield  {author} {\bibinfo {author} {\bibfnamefont {L.}~\bibnamefont
  {Zhang}}, \bibinfo {author} {\bibfnamefont {L.}~\bibnamefont {Zhang}},
  \bibinfo {author} {\bibfnamefont {S.}~\bibnamefont {Niu}}, \ and\ \bibinfo
  {author} {\bibfnamefont {X.-J.}\ \bibnamefont {Liu}},\ }\bibfield  {title}
  {\enquote {\bibinfo {title} {Dynamical classification of topological quantum
  phases},}\ }\href {\doibase 10.1016/j.scib.2018.09.018} {\bibfield  {journal}
  {\bibinfo  {journal} {Sci. Bull.}\ }\textbf {\bibinfo {volume} {63}},\
  \bibinfo {pages} {1385} (\bibinfo {year} {2018})}\BibitemShut {NoStop}%
\bibitem [{\citenamefont {Yu}(2019)}]{Yu2019Hopf}%
  \BibitemOpen
  \bibfield  {author} {\bibinfo {author} {\bibfnamefont {J.}~\bibnamefont
  {Yu}},\ }\bibfield  {title} {\enquote {\bibinfo {title} {{Measuring Hopf
  links and Hopf invariants in a quenched topological Raman lattice}},}\ }\href
  {\doibase 10.1103/PhysRevA.99.043619} {\bibfield  {journal} {\bibinfo
  {journal} {Phys. Rev. A}\ }\textbf {\bibinfo {volume} {99}},\ \bibinfo
  {pages} {043619} (\bibinfo {year} {2019})}\BibitemShut {NoStop}%
\bibitem [{\citenamefont {Zhang}\ \emph
  {et~al.}(2019{\natexlab{a}})\citenamefont {Zhang}, \citenamefont {Zhang},\
  and\ \citenamefont {Liu}}]{Zhang-Zhang-Liu2019_Dyn-Top-Charges}%
  \BibitemOpen
  \bibfield  {author} {\bibinfo {author} {\bibfnamefont {L.}~\bibnamefont
  {Zhang}}, \bibinfo {author} {\bibfnamefont {L.}~\bibnamefont {Zhang}}, \ and\
  \bibinfo {author} {\bibfnamefont {X.-J.}\ \bibnamefont {Liu}},\ }\bibfield
  {title} {\enquote {\bibinfo {title} {Dynamical detection of topological
  charges},}\ }\href {\doibase 10.1103/PhysRevA.99.053606} {\bibfield
  {journal} {\bibinfo  {journal} {Phys. Rev. A}\ }\textbf {\bibinfo {volume}
  {99}},\ \bibinfo {pages} {053606} (\bibinfo {year}
  {2019}{\natexlab{a}})}\BibitemShut {NoStop}%
\bibitem [{\citenamefont {Zhang}\ \emph
  {et~al.}(2019{\natexlab{b}})\citenamefont {Zhang}, \citenamefont {Zhang},
  \citenamefont {Hu}, \citenamefont {Niu},\ and\ \citenamefont
  {Liu}}]{Liu-Int-quench2019}%
  \BibitemOpen
  \bibfield  {author} {\bibinfo {author} {\bibfnamefont {L.}~\bibnamefont
  {Zhang}}, \bibinfo {author} {\bibfnamefont {L.}~\bibnamefont {Zhang}},
  \bibinfo {author} {\bibfnamefont {Y.}~\bibnamefont {Hu}}, \bibinfo {author}
  {\bibfnamefont {S.}~\bibnamefont {Niu}}, \ and\ \bibinfo {author}
  {\bibfnamefont {X.-J.}\ \bibnamefont {Liu}},\ }\bibfield  {title} {\enquote
  {\bibinfo {title} {Emergent topology and symmetry-breaking order in
  correlated quench dynamics},}\ }\href {https://arxiv.org/abs/1903.09144}
  {\bibfield  {journal} {\bibinfo  {journal} {arXiv:1903.09144}\ } (\bibinfo
  {year} {2019}{\natexlab{b}})}\BibitemShut {NoStop}%
\bibitem [{\citenamefont {Zhang}\ \emph
  {et~al.}(2019{\natexlab{c}})\citenamefont {Zhang}, \citenamefont {Zhang},\
  and\ \citenamefont {Liu}}]{Liu-shallow-quench2019}%
  \BibitemOpen
  \bibfield  {author} {\bibinfo {author} {\bibfnamefont {L.}~\bibnamefont
  {Zhang}}, \bibinfo {author} {\bibfnamefont {L.}~\bibnamefont {Zhang}}, \ and\
  \bibinfo {author} {\bibfnamefont {X.-J.}\ \bibnamefont {Liu}},\ }\bibfield
  {title} {\enquote {\bibinfo {title} {Characterizing topological phases by
  quantum quenches: A general theory},}\ }\href
  {https://arxiv.org/abs/1907.08840} {\bibfield  {journal} {\bibinfo  {journal}
  {arXiv:1907.08840}\ } (\bibinfo {year} {2019}{\natexlab{c}})}\BibitemShut
  {NoStop}%
\bibitem [{\citenamefont {\"Unal}\ \emph {et~al.}(2019)\citenamefont {\"Unal},
  \citenamefont {Eckardt},\ and\ \citenamefont {Slager}}]{Unal-Eckardt2019}%
  \BibitemOpen
  \bibfield  {author} {\bibinfo {author} {\bibfnamefont {F.~N.}\ \bibnamefont
  {\"Unal}}, \bibinfo {author} {\bibfnamefont {A.}~\bibnamefont {Eckardt}}, \
  and\ \bibinfo {author} {\bibfnamefont {R.-J.}\ \bibnamefont {Slager}},\
  }\bibfield  {title} {\enquote {\bibinfo {title} {Hopf characterization of
  two-dimensional floquet topological insulators},}\ }\href {\doibase
  10.1103/PhysRevResearch.1.022003} {\bibfield  {journal} {\bibinfo  {journal}
  {Phys. Rev. Research}\ }\textbf {\bibinfo {volume} {1}},\ \bibinfo {pages}
  {022003(R)} (\bibinfo {year} {2019})}\BibitemShut {NoStop}%
\bibitem [{\citenamefont {Fl{\"a}schner}\ \emph {et~al.}(2018)\citenamefont
  {Fl{\"a}schner}, \citenamefont {Vogel}, \citenamefont {Tarnowski},
  \citenamefont {Rem}, \citenamefont {L{\"u}hmann}, \citenamefont {Heyl},
  \citenamefont {Budich}, \citenamefont {Mathey}, \citenamefont {Sengstock},\
  and\ \citenamefont {Weitenberg}}]{Flaschner2018}%
  \BibitemOpen
  \bibfield  {author} {\bibinfo {author} {\bibfnamefont {N.}~\bibnamefont
  {Fl{\"a}schner}}, \bibinfo {author} {\bibfnamefont {D.}~\bibnamefont
  {Vogel}}, \bibinfo {author} {\bibfnamefont {M.}~\bibnamefont {Tarnowski}},
  \bibinfo {author} {\bibfnamefont {B.~S.}\ \bibnamefont {Rem}}, \bibinfo
  {author} {\bibfnamefont {D.-S.}\ \bibnamefont {L{\"u}hmann}}, \bibinfo
  {author} {\bibfnamefont {M.}~\bibnamefont {Heyl}}, \bibinfo {author}
  {\bibfnamefont {J.~C.}\ \bibnamefont {Budich}}, \bibinfo {author}
  {\bibfnamefont {L.}~\bibnamefont {Mathey}}, \bibinfo {author} {\bibfnamefont
  {K.}~\bibnamefont {Sengstock}}, \ and\ \bibinfo {author} {\bibfnamefont
  {C.}~\bibnamefont {Weitenberg}},\ }\bibfield  {title} {\enquote {\bibinfo
  {title} {Observation of dynamical vortices after quenches in a system with
  topology},}\ }\href {\doibase 10.1038/s41567-017-0013-8} {\bibfield
  {journal} {\bibinfo  {journal} {Nat. Phys.}\ }\textbf {\bibinfo {volume}
  {14}},\ \bibinfo {pages} {265} (\bibinfo {year} {2018})}\BibitemShut
  {NoStop}%
\bibitem [{\citenamefont {Tarnowski}\ \emph {et~al.}(2019)\citenamefont
  {Tarnowski}, \citenamefont {{\"{U}}nal}, \citenamefont {Fl{\"{a}}schner},
  \citenamefont {Rem}, \citenamefont {Eckardt}, \citenamefont {Sengstock},\
  and\ \citenamefont {Weitenberg}}]{chernlink}%
  \BibitemOpen
  \bibfield  {author} {\bibinfo {author} {\bibfnamefont {M.}~\bibnamefont
  {Tarnowski}}, \bibinfo {author} {\bibfnamefont {F.~N.}\ \bibnamefont
  {{\"{U}}nal}}, \bibinfo {author} {\bibfnamefont {N.}~\bibnamefont
  {Fl{\"{a}}schner}}, \bibinfo {author} {\bibfnamefont {B.~S.}\ \bibnamefont
  {Rem}}, \bibinfo {author} {\bibfnamefont {A.}~\bibnamefont {Eckardt}},
  \bibinfo {author} {\bibfnamefont {K.}~\bibnamefont {Sengstock}}, \ and\
  \bibinfo {author} {\bibfnamefont {C.}~\bibnamefont {Weitenberg}},\ }\bibfield
   {title} {\enquote {\bibinfo {title} {{Measuring topology from dynamics by
  obtaining the Chern number from a linking number}},}\ }\href {\doibase
  10.1038/s41467-019-09668-y} {\bibfield  {journal} {\bibinfo  {journal} {Nat.
  Commun.}\ }\textbf {\bibinfo {volume} {10}},\ \bibinfo {pages} {1728}
  (\bibinfo {year} {2019})}\BibitemShut {NoStop}%
\bibitem [{\citenamefont {Sun}\ \emph {et~al.}(2018)\citenamefont {Sun},
  \citenamefont {Yi}, \citenamefont {Wang}, \citenamefont {Zhang},
  \citenamefont {Sanders}, \citenamefont {Xu}, \citenamefont {Wang},
  \citenamefont {Schmiedmayer}, \citenamefont {Deng}, \citenamefont {Liu},
  \citenamefont {Chen},\ and\ \citenamefont {Pan}}]{USTC-PKU2018}%
  \BibitemOpen
  \bibfield  {author} {\bibinfo {author} {\bibfnamefont {W.}~\bibnamefont
  {Sun}}, \bibinfo {author} {\bibfnamefont {C.-R.}\ \bibnamefont {Yi}},
  \bibinfo {author} {\bibfnamefont {B.-Z.}\ \bibnamefont {Wang}}, \bibinfo
  {author} {\bibfnamefont {W.-W.}\ \bibnamefont {Zhang}}, \bibinfo {author}
  {\bibfnamefont {B.~C.}\ \bibnamefont {Sanders}}, \bibinfo {author}
  {\bibfnamefont {X.-T.}\ \bibnamefont {Xu}}, \bibinfo {author} {\bibfnamefont
  {Z.-Y.}\ \bibnamefont {Wang}}, \bibinfo {author} {\bibfnamefont
  {J.}~\bibnamefont {Schmiedmayer}}, \bibinfo {author} {\bibfnamefont
  {Y.}~\bibnamefont {Deng}}, \bibinfo {author} {\bibfnamefont {X.-J.}\
  \bibnamefont {Liu}}, \bibinfo {author} {\bibfnamefont {S.}~\bibnamefont
  {Chen}}, \ and\ \bibinfo {author} {\bibfnamefont {J.-W.}\ \bibnamefont
  {Pan}},\ }\bibfield  {title} {\enquote {\bibinfo {title} {Uncover topology by
  quantum quench dynamics},}\ }\href {\doibase 10.1103/PhysRevLett.121.250403}
  {\bibfield  {journal} {\bibinfo  {journal} {Phys. Rev. Lett.}\ }\textbf
  {\bibinfo {volume} {121}},\ \bibinfo {pages} {250403} (\bibinfo {year}
  {2018})}\BibitemShut {NoStop}%
\bibitem [{\citenamefont {Yi}\ \emph {et~al.}(2019{\natexlab{a}})\citenamefont
  {Yi}, \citenamefont {Yu}, \citenamefont {Sun}, \citenamefont {Xu},
  \citenamefont {Chen},\ and\ \citenamefont {Pan}}]{Hopf_fibration2019}%
  \BibitemOpen
  \bibfield  {author} {\bibinfo {author} {\bibfnamefont {C.-R.}\ \bibnamefont
  {Yi}}, \bibinfo {author} {\bibfnamefont {J.-L.}\ \bibnamefont {Yu}}, \bibinfo
  {author} {\bibfnamefont {W.}~\bibnamefont {Sun}}, \bibinfo {author}
  {\bibfnamefont {X.-T.}\ \bibnamefont {Xu}}, \bibinfo {author} {\bibfnamefont
  {S.}~\bibnamefont {Chen}}, \ and\ \bibinfo {author} {\bibfnamefont {J.-W.}\
  \bibnamefont {Pan}},\ }\bibfield  {title} {\enquote {\bibinfo {title}
  {{Observation of the Hopf Links and Hopf Fibration in a 2D Topological Raman
  Lattice}},}\ }\href {http://arxiv.org/abs/1904.11656} {\bibfield  {journal}
  {\bibinfo  {journal} {arXiv:1904.11656}\ } (\bibinfo {year}
  {2019}{\natexlab{a}})}\BibitemShut {NoStop}%
\bibitem [{\citenamefont {Yi}\ \emph {et~al.}(2019{\natexlab{b}})\citenamefont
  {Yi}, \citenamefont {Zhang}, \citenamefont {Zhang}, \citenamefont {Jiao},
  \citenamefont {Cheng}, \citenamefont {Wang}, \citenamefont {Xu},
  \citenamefont {Sun}, \citenamefont {Liu}, \citenamefont {Chen},\ and\
  \citenamefont {Pan}}]{Yi-Shuai2019}%
  \BibitemOpen
  \bibfield  {author} {\bibinfo {author} {\bibfnamefont {C.-R.}\ \bibnamefont
  {Yi}}, \bibinfo {author} {\bibfnamefont {L.}~\bibnamefont {Zhang}}, \bibinfo
  {author} {\bibfnamefont {L.}~\bibnamefont {Zhang}}, \bibinfo {author}
  {\bibfnamefont {R.-H.}\ \bibnamefont {Jiao}}, \bibinfo {author}
  {\bibfnamefont {X.-C.}\ \bibnamefont {Cheng}}, \bibinfo {author}
  {\bibfnamefont {Z.-Y.}\ \bibnamefont {Wang}}, \bibinfo {author}
  {\bibfnamefont {X.-T.}\ \bibnamefont {Xu}}, \bibinfo {author} {\bibfnamefont
  {W.}~\bibnamefont {Sun}}, \bibinfo {author} {\bibfnamefont {X.-J.}\
  \bibnamefont {Liu}}, \bibinfo {author} {\bibfnamefont {S.}~\bibnamefont
  {Chen}}, \ and\ \bibinfo {author} {\bibfnamefont {J.-W.}\ \bibnamefont
  {Pan}},\ }\bibfield  {title} {\enquote {\bibinfo {title} {Observing
  topological charges and dynamical bulk-surface correspondence with ultracold
  atoms},}\ }\href {https://arxiv.org/abs/1905.06478} {\bibfield  {journal}
  {\bibinfo  {journal} {arXiv:1905.06478}\ } (\bibinfo {year}
  {2019}{\natexlab{b}})}\BibitemShut {NoStop}%
\bibitem [{\citenamefont {Gong}\ and\ \citenamefont
  {Ueda}(2018)}]{Gong-Ueda2018}%
  \BibitemOpen
  \bibfield  {author} {\bibinfo {author} {\bibfnamefont {Z.}~\bibnamefont
  {Gong}}\ and\ \bibinfo {author} {\bibfnamefont {M.}~\bibnamefont {Ueda}},\
  }\bibfield  {title} {\enquote {\bibinfo {title} {Topological
  entanglement-spectrum crossing in quench dynamics},}\ }\href {\doibase
  10.1103/PhysRevLett.121.250601} {\bibfield  {journal} {\bibinfo  {journal}
  {Phys. Rev. Lett.}\ }\textbf {\bibinfo {volume} {121}},\ \bibinfo {pages}
  {250601} (\bibinfo {year} {2018})}\BibitemShut {NoStop}%
\bibitem [{\citenamefont {Yang}\ \emph {et~al.}(2018)\citenamefont {Yang},
  \citenamefont {Li},\ and\ \citenamefont {Chen}}]{Yang-Chen2018}%
  \BibitemOpen
  \bibfield  {author} {\bibinfo {author} {\bibfnamefont {C.}~\bibnamefont
  {Yang}}, \bibinfo {author} {\bibfnamefont {L.}~\bibnamefont {Li}}, \ and\
  \bibinfo {author} {\bibfnamefont {S.}~\bibnamefont {Chen}},\ }\bibfield
  {title} {\enquote {\bibinfo {title} {Dynamical topological invariant after a
  quantum quench},}\ }\href {\doibase 10.1103/PhysRevB.97.060304} {\bibfield
  {journal} {\bibinfo  {journal} {Phys. Rev. B}\ }\textbf {\bibinfo {volume}
  {97}},\ \bibinfo {pages} {060304(R)} (\bibinfo {year} {2018})}\BibitemShut
  {NoStop}%
\bibitem [{\citenamefont {McGinley}\ and\ \citenamefont
  {Cooper}(2018)}]{Cooper2018}%
  \BibitemOpen
  \bibfield  {author} {\bibinfo {author} {\bibfnamefont {M.}~\bibnamefont
  {McGinley}}\ and\ \bibinfo {author} {\bibfnamefont {N.~R.}\ \bibnamefont
  {Cooper}},\ }\bibfield  {title} {\enquote {\bibinfo {title} {Topology of
  one-dimensional quantum systems out of equilibrium},}\ }\href {\doibase
  10.1103/PhysRevLett.121.090401} {\bibfield  {journal} {\bibinfo  {journal}
  {Phys. Rev. Lett.}\ }\textbf {\bibinfo {volume} {121}},\ \bibinfo {pages}
  {090401} (\bibinfo {year} {2018})}\BibitemShut {NoStop}%
\bibitem [{\citenamefont {{Qiu}}\ \emph {et~al.}(2019)\citenamefont {{Qiu}},
  \citenamefont {{Deng}}, \citenamefont {{Hu}}, \citenamefont {{Xue}},\ and\
  \citenamefont {{Yi}}}]{YiWei2019}%
  \BibitemOpen
  \bibfield  {author} {\bibinfo {author} {\bibfnamefont {X.}~\bibnamefont
  {{Qiu}}}, \bibinfo {author} {\bibfnamefont {T.-S.}\ \bibnamefont {{Deng}}},
  \bibinfo {author} {\bibfnamefont {Y.}~\bibnamefont {{Hu}}}, \bibinfo {author}
  {\bibfnamefont {P.}~\bibnamefont {{Xue}}}, \ and\ \bibinfo {author}
  {\bibfnamefont {W.}~\bibnamefont {{Yi}}},\ }\bibfield  {title} {\enquote
  {\bibinfo {title} {Fixed points and dynamic topological phenomena in a
  parity-time-symmetric quantum quench},}\ }\href {\doibase
  10.1016/j.isci.2019.09.037} {\bibfield  {journal} {\bibinfo  {journal}
  {iScience}\ }\textbf {\bibinfo {volume} {20}},\ \bibinfo {pages} {392}
  (\bibinfo {year} {2019})}\BibitemShut {NoStop}%
\bibitem [{\citenamefont {Lu}\ and\ \citenamefont {Yu}(2019)}]{Lu2019}%
  \BibitemOpen
  \bibfield  {author} {\bibinfo {author} {\bibfnamefont {S.}~\bibnamefont
  {Lu}}\ and\ \bibinfo {author} {\bibfnamefont {J.}~\bibnamefont {Yu}},\
  }\bibfield  {title} {\enquote {\bibinfo {title} {Stability of
  entanglement-spectrum crossing in quench dynamics of one-dimensional gapped
  free-fermion systems},}\ }\href {\doibase 10.1103/PhysRevA.99.033621}
  {\bibfield  {journal} {\bibinfo  {journal} {Phys. Rev. A}\ }\textbf {\bibinfo
  {volume} {99}},\ \bibinfo {pages} {033621} (\bibinfo {year}
  {2019})}\BibitemShut {NoStop}%
\bibitem [{\citenamefont {Wilczek}\ and\ \citenamefont {Zee}(1983)}]{Azee}%
  \BibitemOpen
  \bibfield  {author} {\bibinfo {author} {\bibfnamefont {F.}~\bibnamefont
  {Wilczek}}\ and\ \bibinfo {author} {\bibfnamefont {A.}~\bibnamefont {Zee}},\
  }\bibfield  {title} {\enquote {\bibinfo {title} {Linking numbers, spin, and
  statistics of solitons},}\ }\href {\doibase 10.1103/PhysRevLett.51.2250}
  {\bibfield  {journal} {\bibinfo  {journal} {Phys. Rev. Lett.}\ }\textbf
  {\bibinfo {volume} {51}},\ \bibinfo {pages} {2250} (\bibinfo {year}
  {1983})}\BibitemShut {NoStop}%
\bibitem [{\citenamefont {Teo}\ and\ \citenamefont
  {Kane}(2010)}]{Teo-Kane2010}%
  \BibitemOpen
  \bibfield  {author} {\bibinfo {author} {\bibfnamefont {J.~C.~Y.}\
  \bibnamefont {Teo}}\ and\ \bibinfo {author} {\bibfnamefont {C.~L.}\
  \bibnamefont {Kane}},\ }\bibfield  {title} {\enquote {\bibinfo {title}
  {Topological defects and gapless modes in insulators and superconductors},}\
  }\href {\doibase 10.1103/PhysRevB.82.115120} {\bibfield  {journal} {\bibinfo
  {journal} {Phys. Rev. B}\ }\textbf {\bibinfo {volume} {82}},\ \bibinfo
  {pages} {115120} (\bibinfo {year} {2010})}\BibitemShut {NoStop}%
\bibitem [{\citenamefont {Barbarino}\ \emph {et~al.}(2019)\citenamefont
  {Barbarino}, \citenamefont {Yu}, \citenamefont {Zoller},\ and\ \citenamefont
  {Budich}}]{Simone-Jinlong-Peter-Jan2019}%
  \BibitemOpen
  \bibfield  {author} {\bibinfo {author} {\bibfnamefont {S.}~\bibnamefont
  {Barbarino}}, \bibinfo {author} {\bibfnamefont {J.}~\bibnamefont {Yu}},
  \bibinfo {author} {\bibfnamefont {P.}~\bibnamefont {Zoller}}, \ and\ \bibinfo
  {author} {\bibfnamefont {J.~C.}\ \bibnamefont {Budich}},\ }\bibfield  {title}
  {\enquote {\bibinfo {title} {{Preparing Atomic Topological Quantum Matter by
  Adiabatic Non-Unitary Dynamics}},}\ }\href {https://arxiv.org/abs/1910.05354}
  {\bibfield  {journal} {\bibinfo  {journal} {arXiv:1910.05354}\ } (\bibinfo
  {year} {2019})}\BibitemShut {NoStop}%
\bibitem [{\citenamefont {Fox}(1948)}]{Fox1948}%
  \BibitemOpen
  \bibfield  {author} {\bibinfo {author} {\bibfnamefont {R.~H.}\ \bibnamefont
  {Fox}},\ }\bibfield  {title} {\enquote {\bibinfo {title} {Homotopy groups and
  torus homotopy groups},}\ }\href {http://www.jstor.org/stable/1969292}
  {\bibfield  {journal} {\bibinfo  {journal} {Ann. Math.}\ }\textbf {\bibinfo
  {volume} {49}},\ \bibinfo {pages} {471} (\bibinfo {year} {1948})}\BibitemShut
  {NoStop}%
\bibitem [{\citenamefont {Pontryagin}(1941)}]{Pontryagin1941}%
  \BibitemOpen
  \bibfield  {author} {\bibinfo {author} {\bibfnamefont {L.~S.}\ \bibnamefont
  {Pontryagin}},\ }\bibfield  {title} {\enquote {\bibinfo {title} {A
  classiﬁcation of mappings of the three-dimensional complex into the
  two-dimensional sphere},}\ }\href {http://mi.mathnet.ru/msb6073} {\bibfield
  {journal} {\bibinfo  {journal} {Mat. Sbornik}\ }\textbf {\bibinfo {volume}
  {9}},\ \bibinfo {pages} {331} (\bibinfo {year} {1941})}\BibitemShut {NoStop}%
\bibitem [{Not()}]{Note_rotate_x}%
  \BibitemOpen
  \href@noop {} {}\bibinfo {note} {We make this basis rotation to increase the
  difference between the initial spin configuration and the ground state of the
  post-quench Hamiltonian. We note that, it can be implemented experimentally
  by imprinting a $\pi/2$ ($-\pi/2$) radio-frequency field pulse along $x$
  direction to the spin states (see, e.g., Ref. \cite{Monroe2017}) before
  (after) the quench process.}\BibitemShut {Stop}%
\bibitem [{\citenamefont {Moore}\ \emph {et~al.}(2008)\citenamefont {Moore},
  \citenamefont {Ran},\ and\ \citenamefont {Wen}}]{Moore-Wen2008}%
  \BibitemOpen
  \bibfield  {author} {\bibinfo {author} {\bibfnamefont {J.~E.}\ \bibnamefont
  {Moore}}, \bibinfo {author} {\bibfnamefont {Y.}~\bibnamefont {Ran}}, \ and\
  \bibinfo {author} {\bibfnamefont {X.-G.}\ \bibnamefont {Wen}},\ }\bibfield
  {title} {\enquote {\bibinfo {title} {Topological surface states in
  three-dimensional magnetic insulators},}\ }\href {\doibase
  10.1103/PhysRevLett.101.186805} {\bibfield  {journal} {\bibinfo  {journal}
  {Phys. Rev. Lett.}\ }\textbf {\bibinfo {volume} {101}},\ \bibinfo {pages}
  {186805} (\bibinfo {year} {2008})}\BibitemShut {NoStop}%
\bibitem [{\citenamefont {Baez}\ and\ \citenamefont
  {Muniain}(1994)}]{Baez1994_Gauge_Knots}%
  \BibitemOpen
  \bibfield  {author} {\bibinfo {author} {\bibfnamefont {J.}~\bibnamefont
  {Baez}}\ and\ \bibinfo {author} {\bibfnamefont {J.~P.}\ \bibnamefont
  {Muniain}},\ }\href {\doibase 10.1142/2324} {\emph {\bibinfo {title} {Gauge
  fields, knots and gravity}}}\ (\bibinfo  {publisher} {World Scientific
  Publishing Company},\ \bibinfo {address} {Singapore},\ \bibinfo {year}
  {1994})\BibitemShut {NoStop}%
\bibitem [{\citenamefont {Hatcher}(2002)}]{Hatcher2005algebraic}%
  \BibitemOpen
  \bibfield  {author} {\bibinfo {author} {\bibfnamefont {A.}~\bibnamefont
  {Hatcher}},\ }\href {https://pi.math.cornell.edu/~hatcher/AT/ATpage.html}
  {\emph {\bibinfo {title} {Algebraic Topology}}}\ (\bibinfo  {publisher}
  {Cambridge University Press},\ \bibinfo {address} {Cambridge, UK},\ \bibinfo
  {year} {2002})\BibitemShut {NoStop}%
\bibitem [{\citenamefont {Yuan}\ \emph {et~al.}(2017)\citenamefont {Yuan},
  \citenamefont {He}, \citenamefont {Wang}, \citenamefont {Deng}, \citenamefont
  {Wang}, \citenamefont {Lian}, \citenamefont {Wang}, \citenamefont {Zhang},
  \citenamefont {Zhang}, \citenamefont {Chang},\ and\ \citenamefont
  {Duan}}]{Duan2017}%
  \BibitemOpen
  \bibfield  {author} {\bibinfo {author} {\bibfnamefont {X.-X.}\ \bibnamefont
  {Yuan}}, \bibinfo {author} {\bibfnamefont {L.}~\bibnamefont {He}}, \bibinfo
  {author} {\bibfnamefont {S.-T.}\ \bibnamefont {Wang}}, \bibinfo {author}
  {\bibfnamefont {D.-L.}\ \bibnamefont {Deng}}, \bibinfo {author}
  {\bibfnamefont {F.}~\bibnamefont {Wang}}, \bibinfo {author} {\bibfnamefont
  {W.-Q.}\ \bibnamefont {Lian}}, \bibinfo {author} {\bibfnamefont
  {X.}~\bibnamefont {Wang}}, \bibinfo {author} {\bibfnamefont {C.-H.}\
  \bibnamefont {Zhang}}, \bibinfo {author} {\bibfnamefont {H.-L.}\ \bibnamefont
  {Zhang}}, \bibinfo {author} {\bibfnamefont {X.-Y.}\ \bibnamefont {Chang}}, \
  and\ \bibinfo {author} {\bibfnamefont {L.-M.}\ \bibnamefont {Duan}},\
  }\bibfield  {title} {\enquote {\bibinfo {title} {{Observation of Topological
  Links Associated with Hopf Insulators in a Solid-State Quantum Simulator}},}\
  }\href {http://stacks.iop.org/0256-307X/34/i=6/a=060302} {\bibfield
  {journal} {\bibinfo  {journal} {Chin. Phys. Lett.}\ }\textbf {\bibinfo
  {volume} {34}},\ \bibinfo {pages} {060302} (\bibinfo {year}
  {2017})}\BibitemShut {NoStop}%
\bibitem [{\citenamefont {Zhang}\ \emph {et~al.}(2017)\citenamefont {Zhang},
  \citenamefont {Pagano}, \citenamefont {Hess}, \citenamefont {Kyprianidis},
  \citenamefont {Becker}, \citenamefont {Kaplan}, \citenamefont {Gorshkov},
  \citenamefont {Gong},\ and\ \citenamefont {Monroe}}]{Monroe2017}%
  \BibitemOpen
  \bibfield  {author} {\bibinfo {author} {\bibfnamefont {J.}~\bibnamefont
  {Zhang}}, \bibinfo {author} {\bibfnamefont {G.}~\bibnamefont {Pagano}},
  \bibinfo {author} {\bibfnamefont {P.~W.}\ \bibnamefont {Hess}}, \bibinfo
  {author} {\bibfnamefont {A.}~\bibnamefont {Kyprianidis}}, \bibinfo {author}
  {\bibfnamefont {P.}~\bibnamefont {Becker}}, \bibinfo {author} {\bibfnamefont
  {H.}~\bibnamefont {Kaplan}}, \bibinfo {author} {\bibfnamefont {A.~V.}\
  \bibnamefont {Gorshkov}}, \bibinfo {author} {\bibfnamefont {Z.~X.}\
  \bibnamefont {Gong}}, \ and\ \bibinfo {author} {\bibfnamefont
  {C.}~\bibnamefont {Monroe}},\ }\bibfield  {title} {\enquote {\bibinfo {title}
  {Observation of a many-body dynamical phase transition with a 53-qubit
  quantum simulator},}\ }\href {http://dx.doi.org/10.1038/nature24654}
  {\bibfield  {journal} {\bibinfo  {journal} {Nature}\ }\textbf {\bibinfo
  {volume} {551}},\ \bibinfo {pages} {601} (\bibinfo {year}
  {2017})}\BibitemShut {NoStop}%
\end{thebibliography}
%

\end{document}